\documentclass[12pt,a4paper]{article}

\usepackage{amsmath}
\usepackage{amssymb}
\usepackage{epsfig}

\def\nostrocostrutto#1\over#2{\mathrel{\mathop{\kern 0pt \rlap 
  {\raise.2ex\hbox{$#1$}}}
  \lower.9ex\hbox{\kern-.190em $#2$}}}
\newcommand{\N}{{\mathcal N}}

\catcode`@=11
\newcount\@tempcntc
\def\@citex[#1]#2{\if@filesw\immediate\write\@auxout{\string\citation{#2}}\fi
  \@tempcnta\z@\@tempcntb\m@ne\def\@citea{}\@cite{\@for\@citeb:=#2\do
    {\@ifundefined
       {b@\@citeb}{\@citeo\@tempcntb\m@ne\@citea\def\@citea{,}{\bf ?}\@warning
       {Citation `\@citeb' on page \thepage \space undefined}}%
    {\setbox\z@\hbox{\global\@tempcntc0\csname b@\@citeb\endcsname\relax}%
     \ifnum\@tempcntc=\z@ \@citeo\@tempcntb\m@ne
       \@citea\def\@citea{,}\hbox{\csname b@\@citeb\endcsname}%
     \else
      \advance\@tempcntb\@ne
      \ifnum\@tempcntb=\@tempcntc
      \else\advance\@tempcntb\m@ne\@citeo
      \@tempcnta\@tempcntc\@tempcntb\@tempcntc\fi\fi}}\@citeo}{#1}}
\def\@citeo{\ifnum\@tempcnta>\@tempcntb\else\@citea\def\@citea{,}%
  \ifnum\@tempcnta=\@tempcntb\the\@tempcnta\else
   {\advance\@tempcnta\@ne\ifnum\@tempcnta=\@tempcntb \else \def\@citea{--}\fi
    \advance\@tempcnta\m@ne\the\@tempcnta\@citea\the\@tempcntb}\fi\fi}
\catcode`@=12
\begin{document}

%****************************************************************

% Titlepage

%****************************************************************

\setcounter{page}{0}
\thispagestyle{empty}
\begin{titlepage}

% BUTP Nr.
\vspace*{-1cm}
\hfill \parbox{3.5cm}{BUTP-2001/05 \\ 
hep-ph/0102249 \\
20. February , 2001
%hep-ph/0102249
}   
\vfill

% Title
\begin{center}
  {\large {\bf
On the quest for unification - simplicity and antisimplicity.}
      \footnote{
      Dedicated to Alberto Sirlin on the occasion of his birth day.}  }
\vfill
% Authors

{\bf
    Peter Minkowski } \\
    Institute for Theoretical Physics \\
    University of Bern \\
    CH - 3012 Bern, Switzerland
%   \vspace*{0.5cm} \\  

\end{center}

\vfill

\begin{abstract}
The road towards unification of elementary interactions is thought to
start on the solid ground of a universal local gauge principle.
I discuss the different types of bosonic gauge symmetries in gravitational and
nongravitational (standard model) interactions and their extensions both
fermionic, bosonic and with respect to space-time dimensions. 
The apparently paradoxical
size and nature of the cosmological constant is sketched, which at first sight
does not readily yield a clue as to
the envelopping symmetry structure of a unified theory.
Nevertheless a tentative outlook is given encouraging to proceed on this road.
\end{abstract}

%\vspace{-15cm}

%\rightline{MPI-PhT/97-46}
%\rightline{hep-ph/9707393} 
%\rightline{January 22, 2000}

\vfill
\end{titlepage}

%****************************************************************

\newpage
%\noindent
{\bf In collaboration with \\
Sonja Kabana, Wolfgang Ochs and Luzi Bergamin.}

\section*{Topics :}

\noindent
\begin{displaymath}
\begin{array}{llr}
 & & \mbox{length scale} \\
 \hline \\
1) & \mbox{
\begin{tabular}[t]{l}
the ways of gravitation \\
gauging spin (- Lie group) \\
$e^{\ a}_{\ \mu} \ \leftrightarrow \ g_{\ \mu \nu}$
\end{tabular}}
 & \lambda \ > \ 
 \begin{array}{c}
1 
 \vspace*{0.3cm} \\
\hline	\vspace*{-0.3cm} \\
2 \ eV 
\end{array}
\ \sim \ 10^{\ 8} \ fm \\
\multicolumn{3}{l}
{\mbox{(no) response to vacuum energy density}} 
\vspace*{0.3cm} \\
2) & + \ QED \ \mbox{
\begin{tabular}[t]{l}
gauging charge $U1_{\ em}$ \\
$\gamma, \ e^{\ \pm}, \ \mu^{\ \pm}, \ \tau^{\ \pm}$ 
\end{tabular}}
 & \lambda \ > \ \lambda^{e}_{cl} \ =
 \begin{array}{c}
\alpha 
 \vspace*{0.3cm} \\
\hline	\vspace*{-0.3cm} \\
m_{\ e} 
\end{array}
\ \sim \ 3 \ fm \\
\multicolumn{3}{l}
{\mbox{what is the relation between charge and spin ?}} 
\vspace*{0.3cm} \\
3) & + \ QCD \mbox{\begin{tabular}[t]{l} $SU3_{\ c}$ \\
gauging color charges \\
$gl^{\ A}, \ ( \ u,d;c,s;t,b \ )^{ c}$ \\
\hspace*{1.0cm} and $ q \ \rightarrow \ \overline{q}$
\end{tabular}}
 & \lambda \ > \ ( \sqrt{2} G_{F} )^{1/2} \ \sim \ 10^{-3} \ fm \\
4) & + \ SU2_{L} \times U1_{Y} \ \mbox{\begin{tabular}[t]{l} the electroweak \\
completion
\end{tabular}}
 & \lambda \ > 
\ \begin{array}{c}
\alpha 
 \vspace*{0.3cm} \\
\hline	\vspace*{-0.3cm} \\
\pi  
\end{array}
 \ ( \sqrt{2} G_{F} )^{1/2} \ \sim \ 10^{-6} \ fm \\
\multicolumn{3}{l}
{\mbox{gauge groups, + exact , - broken}} \\
\multicolumn{3}{l}
{\begin{array}{ccc ccc ccc}
& SO \ (3,1) & \times & SU3_{c} & \times & U1_{em} \ \rightarrow & SU2_{L} \times U1_{Y} & & \\
& +          & & +              &  & +                              & - & & \\  
& gr         & & gl             & & \gamma                       & W^{\pm}, Z &  \varphi_{H} & \\
\# \ d.o.f. & 2 & & 16            & &  2                           & 9          & 1 & \rightarrow \ 30
\end{array}}
\end{array}
\end{displaymath}

\newpage

{\bf topics continued}

\begin{displaymath}
\begin{array}{llr}
 & & \mbox{length scale} \\
  \hline \\
5) & \mbox{\begin{tabular}[t]{l} unification of charges \\
$SU3_{c} \times SU2_{L} \times U1_{Y} \ \rightarrow \ SO10 \ \rightarrow \ E6 \cdots$ \\
\end{tabular}}
 & \lambda \ \sim \ 10^{-17} \ fm \\
6) & \mbox{\begin{tabular}[t]{l}  gauging fermionic charges \\
susy $\leftrightarrow$ anti-simplicity
\end{tabular}}
 & \lambda \ \sim \ 10^{-4} \ fm \\
7) & \mbox{\begin{tabular}[t]{ll} exact & broken symmetries \\
+ gauge symmetries, CPT & leptonic numbers \\
			& $L_{e,\mu,\tau} \leftrightarrow ( \ \nu \ , \ \cal{N} \ )$ \\
			& neutrino pairs 
\end{tabular}}
  & \lambda \ \sim \ 10^{-11} \ fm  \\
8) & \mbox{\begin{tabular}[t]{l} primary breakdown \\
difficulties
\end{tabular}}
 & \lambda \ \sim \ 10^{-17} \ fm \\
9) & \mbox{\begin{tabular}[t]{l} 4 dimensions and 
cosmological term : 
a paradox
\end{tabular}}
  & \mbox{all}  \\
10) & \mbox{\begin{tabular}[t]{l} outlook ( quo vadis ? )
\end{tabular}}
\end{array}
\end{displaymath}

\section{The ways of gravitation}

\noindent
The base quantities involve the structural forms with respect to tangent space
of vierbein, spin connection and Riemann curvature

\begin{equation}
\begin{array}{l}
\label{eq:1}
e^{\ a} \ = 
\ \left ( \ e_{\ 1} \ \right )^{ a} \ = \ d x^{\mu} \ e^{ a}_{\ \mu}
\hspace*{0.2cm} ; \hspace*{0.2cm}
\omega^{\ a}_{\ \ b}
\ = \ \left ( \ \omega_{\ 1} \ \right )^{ a}_{\ b} \ = \ d x^{\mu} 
\ \left ( \omega_{\ \mu} \right )^{ a}_{\ b}
%\hspace*{0.2cm} ; \hspace*{0.2cm}
\vspace*{0.3cm} \\
R^{\ a}_{\ \ b}
\ = \ \left ( \ R_{\ 2} \ \right )^{ a}_{\ b} \ = \ \frac{1}{2} \ d x^{\mu} \ \wedge \ d x^{\nu}
\ \left ( R_{\ \mu \nu} \right )^{ a}_{\ b}
\vspace*{0.3cm} \\
g_{\ \mu \nu} \ = \ e^{ a}_{\ \mu} \ \eta_{\ a b} \ e^{ b}_{\ \nu}
\hspace*{0.2cm} ; \hspace*{0.2cm}
\left ( \ \Gamma_{ \mu} \ \right )^{\nu}_{\ \varrho}
\ = \ e^{\nu}_{\ a} \ \left \lbrack \ \partial_{ \mu} \ e^{a}_{\ \varrho}
\ + \ \left ( \ \omega_{ \mu} \ \right )^{a}_{\ b} \ e^{b}_{\ \varrho}
\ \right \rbrack
\vspace*{0.3cm} \\
R^{\varrho}_{\ \sigma \ \mu \nu} 
\ =
\ e^{ \varrho}_{\ a}
\ e^{b}_{\ \sigma}
\ \left ( R_{\ \mu \nu} \right )^{ a}_{\ b}
\end{array}
\end{equation}

\noindent
with the structural relations

\begin{equation}
\begin{array}{l}
\label{eq:2}
d \ e^{\ a}
\ + \ 
\omega^{\ a}_{\ b} \ e^{\ b} 
\ = \ 2 \ T^{\ a}
\ = \ 2 \ \left ( \ T_{\ 2} \ \right )^{ a}
%\hspace*{0.2cm} ; \hspace*{0.2cm}
\vspace*{0.3cm} \\
T^{\ a} 
\ = \ \frac{1}{2} \ d x^{\mu} \ \wedge \ d x^{\varrho} \ T^{\ a}_{ \mu \ \ \varrho}
\vspace*{0.3cm} \\
%\hspace*{0.2cm} ; \hspace*{0.2cm}
T^{\ \sigma}_{ \left \lbrack \mu \ \ \varrho \right \rbrack} \ = 
\ e^{\sigma}_{\ a} \ T^{\ a}_{ \mu \ \ \varrho}
\ =  \ \frac{1}{2} \ \left \lbrack \left ( \ \Gamma_{ \mu} \ \right )^{\sigma}_{\ \varrho}
\ - \ \left ( \ \Gamma_{ \varrho} \ \right )^{\sigma}_{\ \mu} \ \right \rbrack
\vspace*{0.3cm} \\
R^{\ a}_{\ \ b} \ = \ - \ \left ( \ d \ \omega \ + \ \omega^{\ 2} \ \right )^{\ a}_{\ \ b}
\end{array}
\end{equation}

\noindent
In eq. (\ref{eq:1}) $g_{\ \mu \nu}$ and $\Gamma^{\ \nu}_{\ \mu \ \varrho}$
denote the metric and vector connection respectively. $\eta_{\ a b}$ denotes
the tangent space Lorentz invariant metric.
Vector and spin connections are characterized by their minimal representatives

\begin{equation}
\begin{array}{l}
\label{eq:3}
\Gamma^{\ \ \sigma}_{\ \mu \ \ \varrho} \ =
\ = \ \left ( \ \Gamma_{ \mu} \ \right )^{\sigma}_{\ \varrho} 
\hspace*{0.2cm} ; \hspace*{0.2cm}
T^{\ \sigma}_{ \left \lbrack \mu \ \ \varrho \right \rbrack} \ = 
\ T^{\ \sigma}_{ \mu \ \ \varrho} \ = 
\ g^{\ \sigma \tau}
\ T_{\ \left \lbrack \mu \varrho \right \rbrack \ ; \ \tau}
\vspace*{0.3cm} \\
\gamma^{\ \ \sigma}_{\ \left \lbrace \mu \ \ \varrho \right \rbrace} \ =
\ g^{\ \sigma \tau}
\ \gamma_{\ \left \lbrace \mu \varrho \right \rbrace \ ; \ \tau}
\hspace*{0.2cm} ; \hspace*{0.2cm}
\gamma_{\ \left \lbrace \mu \varrho \right \rbrace \ ; \ \sigma}
\ = \ \frac{1}{2} \ \left ( 
\ \partial_{\mu} g_{\varrho \sigma}
\ + \ \partial_{\varrho} g_{\mu \sigma}
\ - \ \partial_{\sigma} g_{\mu \varrho} \ \right )
\vspace*{0.3cm} \\
\hat{\omega}
\hspace*{0.2cm} : \hspace*{0.2cm}
d \ e^{\ a} \ + \ \hat{\omega}^{\ a}_{\ \ b} \ e^{\ b} \ = \ 0
\vspace*{0.3cm} \\
\hat{\omega}_{\ \mu}^{\ \left \lbrack a \ b \right \rbrack} = \ 
\ \eta^{\ b b'} \ \left ( \ \hat{\omega}_{\ \mu} \ \right )^{\ a}_{\ \ b'}
\ = \ \frac{1}{2} 
\ \left \lbrace
\begin{array}{l}
e^{ \ \varrho b} \ \left ( 
\ \partial_{\ \varrho} \ e^{\ a}_{\ \mu}
\ - \ \partial_{\ \mu} \ e^{\ a}_{\ \varrho}
\ \right )
\vspace*{0.3cm} \\
- \ \left \lbrack \ a \ \leftrightarrow \ b \ \right \rbrack
\vspace*{0.3cm} \\
+ \ e^{\ \varrho b} \ e^{\ \sigma a} \ e_{\ \mu c}
\ \left ( 
\ \partial_{\ \varrho} \ e^{\ c}_{\ \sigma}
\ - \ \partial_{\ \sigma} \ e^{\ c}_{\ \varrho}
\ \right )
\end{array}
\right .
\end{array}
\end{equation}

\noindent
In eqs. (\ref{eq:2}) and (\ref{eq:3}) and in the following tensor (or tangent space)
indices in brackets denote antisymmetric $\left \lbrack \right \rbrack$ and
symmetric $\left \lbrace \right \rbrace$ symmetric combinations respectively.

\noindent
{\bf Torsion and contorsion}
\vspace*{0.1cm}

\noindent
The difference of two connections, for a given metric, defines a covariant quantity, a three tensor
in the case of vector or spin connections, with the respective relations 

\begin{equation}
\begin{array}{l}
\label{eq:4}
\omega_{\ \mu \ ; \ \left \lbrack \nu \varrho \right \rbrack} \ =
\ e_{\ \nu a} \ e_{\ \varrho}^{\ \ b} 
\ \left ( \omega_{\ \mu} \right )^{ a}_{\ b}
\hspace*{0.2cm} ; \hspace*{0.2cm}
\Gamma_{\ \mu \varrho \ ; \ \nu} \ =
\ g_{\ \nu \sigma} \ \Gamma^{\ \ \sigma}_{\ \mu \ \ \varrho} 
\vspace*{0.3cm} \\
\Gamma_{\ \mu \varrho \ ; \ \nu} \ =
\ e_{\ \nu a} \ \partial_{\ \mu} \ e^{\ a}_{\ \varrho} \ + 
\ \omega_{\ \mu \ ; \ \left \lbrack \nu \varrho \right \rbrack}
\ = 
\ \gamma_{\ \left \lbrace \mu \varrho \right \rbrace \ ; \ \nu}
\ + \ \Delta \Gamma_{\ \mu \varrho \ ; \ \nu} 
\vspace*{0.3cm} \\
\omega_{\ \mu \ ; \ \left \lbrack \nu \varrho \right \rbrack} \ =
\ \hat{\omega}_{\ \mu \ ; \ \left \lbrack \nu \varrho \right \rbrack} \ +
\ \Delta \omega_{\ \mu \ ; \ \left \lbrack \nu \varrho \right \rbrack} 
\end{array}
\end{equation}

\noindent
and

\begin{equation}
\begin{array}{l}
\label{eq:5}
\gamma_{\ \left \lbrace \mu \varrho \right \rbrace \ ; \ \nu}
\ =
\ e_{\ \nu a} \ \partial_{\ \mu} \ e^{\ a}_{\ \varrho} \ + 
\ \hat{\omega}_{\ \mu \ ; \ \left \lbrack \nu \varrho \right \rbrack}
\vspace*{0.3cm} \\
%\hspace*{0.2cm} ; \hspace*{0.2cm}
\hat{\omega}_{\ \mu \ ; \ \left \lbrack \nu \varrho \right \rbrack} \ =
\ \frac{1}{2} 
\ \left \lbrace
\begin{array}{l}
\ e_{\ \nu a} 
\ \left ( \ \partial_{\ \varrho} \ e_{\ \mu}^{\ a}
\ - \ \partial_{\ \mu} \ e_{\ \varrho}^{\ a}
\ \right ) 
\vspace*{0.3cm} \\
\ -
\ e_{\ \varrho a} 
\ \left ( \ \partial_{\ \nu} \ e_{\ \mu}^{\ a}
\ - \ \partial_{\ \mu} \ e_{\ \nu}^{\ a}
\ \right )
\vspace*{0.3cm} \\
\ +
\ e_{\ \mu a} 
\ \left ( \ \partial_{\ \varrho} \ e_{\ \nu}^{\ a}
\ - \ \partial_{\ \nu} \ e_{\ \varrho}^{\ a}
\ \right )
\end{array}
\ \right \rbrace
\vspace*{0.3cm} \\
\mbox{contorsion :}
\ \Delta \omega_{\ \mu \ ; \ \left \lbrack \nu \varrho \right \rbrack} 
\ \equiv \ K_{\ \mu \ ; \ \left \lbrack \nu \varrho \right \rbrack} 
\hspace*{0.2cm} ; \hspace*{0.2cm}
\Delta \Gamma_{\ \mu \varrho \ ; \ \nu} \ =
\ K_{\ \mu \ ; \ \left \lbrack \nu \varrho \right \rbrack} 
\vspace*{0.3cm} \\
T_{\ \left \lbrack \mu \varrho \right \rbrack \ ; \ \nu} \ =
\ \frac{1}{2} 
\ \left (
\ \Delta \Gamma_{\ \mu \varrho \ ; \ \nu} \ -
\ \Delta \Gamma_{\ \varrho \mu \ ; \ \nu}
\ \right )
\ =
\ \frac{1}{2} 
\ \left (
\ K_{\ \mu \ ; \ \left \lbrack \nu \varrho \right \rbrack} 
\ -
\ K_{\ \varrho \ ; \ \left \lbrack \nu \mu \right \rbrack} 
\ \right )
\vspace*{0.3cm} \\
K_{\ \mu \ ; \ \left \lbrack \nu \varrho \right \rbrack} 
\ =
\ T_{\ \left \lbrack \nu \varrho \right \rbrack \ ; \ \mu} \ -
\ T_{\ \left \lbrack \mu \nu \right \rbrack \ ; \ \varrho} \ -
\ T_{\ \left \lbrack \varrho \mu \right \rbrack \ ; \ \nu}
%K_{\ \nu \ ; \ \left \lbrack \mu \varrho \right \rbrack} 
%\ =
%\ T_{\ \left \lbrack \mu \varrho \right \rbrack \ ; \ \nu} \ -
%\ T_{\ \left \lbrack \nu \mu \right \rbrack \ ; \ \varrho} \ -
%\ T_{\ \left \lbrack \varrho \nu \right \rbrack \ ; \ \mu}
\end{array}
\end{equation}

\noindent
It follows that all covariant quantities derived from the
difference of vector or spin connections and their minimal forms are
equivalently determined from either the torsion or contorsion three tensor

\begin{equation}
\begin{array}{l}
\label{eq:6}
\Delta \Gamma_{\ \left \lbrace \mu \varrho \right \rbrace \ ; \ \nu} \ =
 \begin{array}[t]{l}
\frac{1}{2}
\ \left (
\Delta \Gamma_{\ \mu \varrho \ ; \ \nu} \ +
\ \Delta \Gamma_{\ \varrho \mu \ ; \ \nu}
\ \right )
\ =
\ \frac{1}{2}
\ \left (
\ K_{\ \mu \ ; \ \left \lbrack \nu \varrho \right \rbrack} 
\ +
\ K_{\ \varrho \ ; \ \left \lbrack \nu \mu \right \rbrack} 
\ \right )
\vspace*{0.3cm} \\
\ =
\ T_{\ \left \lbrack \nu \varrho \right \rbrack \ ; \ \mu} \ +
\ T_{\ \left \lbrack \nu \mu \right \rbrack \ ; \ \varrho}
\end{array}
\end{array}
\end{equation}

\noindent
A special case arises if torsion and contorsion tensors are totally antisymmetric

\begin{equation}
\begin{array}{l}
\label{eq:7}
\Delta \Gamma_{\ \left \lbrace \mu \varrho \right \rbrace \ ; \ \nu} \ = \ 0 
\hspace*{0.2cm} \rightarrow \hspace*{0.2cm}
%\vspace*{0.3cm} \\
T_{\ \left \lbrack \mu \varrho \right \rbrack \ ; \ \nu} \ =
\ T_{\ \left \lbrack \mu \varrho \nu \right \rbrack} \ =
\ K_{\ \mu \ ; \ \left \lbrack \nu \varrho \right \rbrack} \ =
\ K_{\ \left \lbrack \mu \nu \varrho \right \rbrack}
\end{array}
\end{equation}

\noindent
Then autoparallel curves and geodesics, obeying 
inequivalent differential equations in general, coincide

\begin{equation}
\begin{array}{l}
\label{eq:8}
\ddot{Y}^{\ \nu} \ + \ \Gamma_{\ \mu \ \ \varrho}^{\ \ \nu} 
\ \dot{Y}^{\ \mu} \ \dot{Y}^{\ \varrho}
\ = \ 0
\hspace*{0.4cm} \leftrightarrow \hspace*{0.4cm}
\ddot{X}^{\ \nu} \ + \ \gamma_{\ \left \lbrace \mu \varrho \right \rbrace}^{\ \nu} 
\ \dot{X}^{\ \mu} \ \dot{X}^{\ \varrho}
\ = \ 0
\end{array}
\end{equation}

\noindent
{\bf Action densities}

\noindent
We will restrict the discussion to minimal connections in the following,
because torsion (contorsion) fields become dynamically dependent on matter fields,
through their coupling to gravity. This is of course particularly so in supergravity
theories where gravitinos give rise to torsion 

\begin{equation}
\begin{array}{l}
\label{eq:9}
T_{\ \left \lbrack \mu \varrho \right \rbrack}^{\ a} \ \propto
\ \sum_{\ i} \ \overline{\psi}_{\ \mu}^{\ i} \ \gamma^{\ a} \ \psi_{\ \varrho}^{\ i}
\end{array}
\end{equation}

\noindent
In eq. (\ref{eq:9}) $\psi_{\ \mu}^{\ i}$ $i \ = \ 1 \cdots N$ denote N irreducible spin 3/2
fields, obeying in $d \ = \ 4$ dimensions a Majorana condition 
$C \ \gamma_{\ 0} \ \left ( \ \psi_{\ \mu}^{\ i} \ \right )^{\ \dagger} \ = \ \psi_{\ \mu}^{\ i}$ .

\noindent
Let us consider a class of action densities of the Brans-Dicke type \cite{BraDi}
\vspace*{0.1cm}

\begin{equation}
\begin{array}{l}
\label{eq:10}
s \ = \ e \ Q \ R \ + \ e \ {\cal{L}}_{\ matter}
\vspace*{0.3cm} \\
Q \ ( \ x \ ) \ \rightarrow
\ \left \lbrace 
\ \begin{array}{c}
1
 \vspace*{0.3cm} \\
\hline	\vspace*{-0.3cm} \\
16 \pi \ G_{\ N}
\end{array}
%\vspace*{0.3cm} \\
\ = 
\ \begin{array}{c}
1
 \vspace*{0.3cm} \\
\hline	\vspace*{-0.3cm} \\
16 \pi \ l_{\ Pl}^{\ 2}
\end{array}
\right .
\hspace*{0.3cm} \mbox{for} \hspace*{0.3cm}
\ \begin{array}{l}
 d \ \rightarrow \ 4
 \vspace*{0.3cm} \\
 \mbox{\begin{tabular}{l} in the long range \\
 classical limit \end{tabular}}
\end{array}
 \vspace*{0.3cm} \\
e \ = \ det \ ( \ e_{\ \mu}^{\ a} \ )
\hspace*{0.3cm} ; \hspace*{0.3cm}
l_{\ Pl} \ = \ 1.62 \ 10^{\ -20} \ fm
\end{array}
\end{equation}

\noindent
In eq. (\ref{eq:10}) Q is thought to be a quadratic function
of a set of real scalar fields

\begin{equation}
\begin{array}{l}
\label{eq:11}
Q \ = \ \frac{1}{2} \ C^{\ \alpha \beta} \ \varphi_{\ \alpha} \ ( \ x \ )
\ \varphi_{\ \beta} \ ( \ x \ )
\end{array}
\end{equation}

\noindent
A nontrivial set of scalars $\varphi_{\ \alpha}$ is necessary to generate
simultaneously and spontaneously cuvature as well as the Newton constant. 

\noindent
Infinitesimal variation of s with respect to e or g then yields

\begin{equation}
\begin{array}{l}
\label{eq:12}
\delta \ s \ = \ e \ \left ( \ \delta \ g^{\ \mu \nu} \ \right ) \ s_{\ \mu \nu}
\hspace*{0.3cm} ; \hspace*{0.3cm}
%\vspace*{0.3cm} \\
s_{\ \mu \nu} \ =
\ \left \lbrace
\begin{array}{l}
\ \left \lbrack \ D_{\ \mu} \ D_{\ \nu} \ - \ g_{\ \mu \nu} \ D^{\ 2} \ \right \rbrack
\ Q 
\vspace*{0.3cm} \\
+ \ Q \ E_{\ \mu \nu} \ + \ \frac{1}{2} \ \vartheta_{\ \mu \nu}^{\ matter}
\end{array}
\right .
\vspace*{0.3cm} \\
E_{\ \mu \nu} \ = \ R_{\ \mu \nu} \ - \ \frac{1}{2} \ g_{\ \mu \nu} \ R
\end{array}
\end{equation}

\noindent
In eq. (\ref{eq:12}) covariant derivatives relate to the minimal (metric) connection.
$E_{\ \mu \nu}$, $R_{\ \mu \nu}$ denote Einstein and Ricci tensor respectively
and R the curvature scalar. The Riemann curvature tensor takes its conventional form

\begin{equation}
\begin{array}{l}
\label{eq:14}
R^{\ \mu}_{\ \ \nu \ \left \lbrack \sigma \tau \right \rbrack} \ =
\ - \ g^{\ \mu \mu'}  
\ \left (
 \begin{array}{c}
 \gamma_{\ \left \lbrace \mu' \tau \right \rbrace \ ; \ \alpha}
\ g^{\ \alpha \beta}
\ \gamma_{\ \left \lbrace \sigma  \nu \right \rbrace \ ; \ \beta}
\ - \ \left \lbrack \ \sigma \ \leftrightarrow \ \tau \ \right \rbrack
\vspace*{0.3cm} \\
- \ \partial_{\ \tau}
\ \gamma_{\ \left \lbrace \sigma \nu \right \rbrace \ ; \ \mu'}
+ \ \partial_{\ \sigma}
\ \gamma_{\ \left \lbrace \tau \nu \right \rbrace \ ; \ \mu'}
\end{array}
\ \right )
\end{array}
\end{equation}

\noindent
It is instructive to retain the characteristic part $\chi$ of the Riemann curvature tensor
upon dropping terms containing (quadratically) single space time derivatives

\begin{equation}
\begin{array}{l}
\label{eq:15}
\chi^{\ \mu}_{\ \ \nu \ \left \lbrack \sigma \tau \right \rbrack} \ =
\ \frac{1}{2}
\ g^{\ \mu \mu'}  
\ \left (
 \begin{array}{c}
 \partial_{\ \tau} \ \partial_{\ \nu} \ g_{\ \sigma \mu'}
\ - \ \partial_{\ \tau} \ \partial_{\ \mu'} \ g_{\ \sigma \nu}
\vspace*{0.3cm} \\
 -  \ \partial_{\ \sigma} \ \partial_{\ \nu} \ g_{\ \tau \mu'}
\ + \ \partial_{\ \sigma} \ \partial_{\ \mu'} \ g_{\ \tau \nu}
\end{array}
\ \right )
\end{array}
\end{equation}

\noindent
For definiteness we exhibit the structure of the Ricci tensor and the
curvature scalar explicitely

\begin{equation}
\begin{array}{l}
\label{eq:16}
R_{\ \nu \tau} \ = 
\left \lbrace
 \begin{array}{c}
 - \ g^{\ \alpha' \beta'}  
\ g^{\ \alpha \beta}
 \left (
 \begin{array}{c}
 \gamma_{\ \left \lbrace \alpha' \tau \right \rbrace \ ; \ \alpha}
\ \gamma_{\ \left \lbrace \beta' \nu \right \rbrace \ ; \ \beta}
\ - 
\ \gamma_{\ \left \lbrace \alpha' \beta' \right \rbrace \ ; \ \alpha}
\ \gamma_{\ \left \lbrace \tau \nu \right \rbrace \ ; \ \beta}
\end{array}
 \right )
\vspace*{0.3cm} \\
+ \ 
\ g^{\ \alpha \beta}
\ \left (
\ \partial_{\ \tau} 
\ \gamma_{\ \left \lbrace \alpha \nu \right \rbrace \ ; \ \beta}
\ - 
\ \partial_{\ \alpha} 
\ \gamma_{\ \left \lbrace \tau \ \nu \right \rbrace \ ; \ \beta}
\ \right )
\end{array}
\right \rbrace
\vspace*{0.3cm} \\
\chi_{\ \nu \tau} \ =
\ \frac{1}{2}
\ g^{\ \alpha \beta}  
\ \left (
 \begin{array}{c}
 \partial_{\ \tau} \ \partial_{\ \nu} \ g_{\ \alpha \beta}
\ - \ \partial_{\ \tau} \ \partial_{\ \alpha} \ g_{\ \beta \nu}
\vspace*{0.4cm} \\
 -  \ \partial_{\ \nu} \ \partial_{\ \alpha} \ g_{\ \beta \tau}
\ + \ \partial_{\ \alpha} \ \partial_{\ \beta} \ g_{\ \tau \nu}
\end{array}
 \right )
 \ = D_{\ \nu \tau}^{\ \ \nu' \tau'} \ g_{\ \nu' \tau'}
\vspace*{0.4cm} \\
D_{\ \nu \tau}^{\ \ \nu' \tau'} \ ( \ g \ ; \ \partial_{\ \varrho} \ )\ =
\ \frac{1}{2}
\ \left (
 \begin{array}{c}
\ \delta_{\ \nu}^{\ \nu'}
\ \delta_{\ \tau}^{\ \tau'}
\ g^{\ \alpha \beta}
\ \partial_{\ \alpha} \ \partial_{\ \beta} 
\ +
\ g^{\ \nu' \tau'}
\ \partial_{\ \nu} \ \partial_{\ \tau} 
\vspace*{0.3cm} \\
- \ \delta_{\ \nu}^{\ \nu'} 
\ g^{\ \tau' \alpha}
\ \partial_{\ \tau} \ \partial_{\ \alpha} 
- \ \delta_{\ \tau}^{\ \tau'} 
\ g^{\ \nu' \alpha}
\ \partial_{\ \nu} \ \partial_{\ \alpha} 
\end{array}
 \right )
\end{array}
\end{equation}

\noindent
The second order differential operator $D_{\ \nu \tau}^{\ \ \nu' \tau'}$ is understood
to by symmetrized with respect to the index pair $^{\nu' \tau'}$.

\noindent
Finally the curvature scalar becomes

\begin{equation}
\begin{array}{l}
\label{eq:17}
R \ = 
\left \lbrace
 \begin{array}{c}
 - \ g^{\ \alpha' \beta'}  
\ g^{\ \alpha'' \beta''}  
\ g^{\ \alpha \beta}
 \left (
 \begin{array}{c}
 \gamma_{\ \left \lbrace \alpha' \alpha'' \right \rbrace \ ; \ \alpha}
\ \gamma_{\ \left \lbrace \beta' \beta'' \right \rbrace \ ; \ \beta}
\vspace*{0.3cm} \\
\hspace*{0.5cm} - 
\ \gamma_{\ \left \lbrace \alpha' \beta' \right \rbrace \ ; \ \alpha}
\ \gamma_{\ \left \lbrace \alpha'' \beta'' \right \rbrace \ ; \ \beta}
\end{array}
 \right )
\vspace*{0.3cm} \\
+ \ 
\ g^{\ \alpha \beta}
\ g^{\ \alpha' \beta'}  
\ \left (
\ \partial_{\ \alpha'} 
\ \gamma_{\ \left \lbrace \alpha \beta' \right \rbrace \ ; \ \beta}
\ - 
\ \partial_{\ \alpha} 
\ \gamma_{\ \left \lbrace \alpha' \beta' \right \rbrace \ ; \ \beta}
\ \right )
\end{array}
\right \rbrace
\vspace*{0.5cm} \\
\chi \ =
\ g^{\ \alpha \beta}
\ \ g^{\ \alpha' \beta'}  
 \left (
 \begin{array}{c}
\ \partial_{\ \alpha'} 
\ \partial_{\ \beta'} 
\ g_{\ \alpha \beta}
\ - \ \partial_{\ \alpha'} 
\ \partial_{\ \alpha} 
\ g_{\ \beta \beta'}
\end{array}
 \right )
\ = \ D^{\ \nu' \tau'} \ g_{\ \nu' \tau'}
\vspace*{0.5cm} \\
D^{\ \nu' \tau'} \ = 
\ \left (
\ g^{\ \nu' \tau' }
\ g^{\ \alpha \beta}
\ -
\ g^{\ \nu' \alpha}
\ g^{\ \tau' \beta}
\ \right )
\ \partial_{\ \alpha} 
\ \partial_{\ \beta } 
\end{array}
\end{equation}

\noindent
We give also the form of the action density s containing only first derivatives
in quadratic form modulo a total divergence

\begin{equation}
\begin{array}{l}
\label{eq:17a}
s \ = 
\ \left \lbrace
\ \begin{array}{l}
e \ Q 
\ g^{\ \alpha \beta}
\ \left (
\ - \ \gamma_{\ \left \lbrace \alpha \sigma \right \rbrace}^{\ \varrho} 
\ \gamma_{\ \left \lbrace \beta \varrho \right \rbrace}^{\ \sigma} 
\ + \ \gamma_{\ \left \lbrace \alpha \beta \right \rbrace}^{\ \varrho} 
\ \gamma_{\ \left \lbrace \varrho \sigma \right \rbrace}^{\ \sigma} 
\ \right )
\vspace*{0.3cm} \\
+ \ e \ \left ( \ \partial_{\ \alpha} \ Q \ \right )
\ \left (
\ g^{\ \varrho \sigma}
\ \gamma_{\ \left \lbrace \varrho \sigma \right \rbrace}^{\ \alpha} 
\ - 
\ g^{\ \alpha \beta}
\ \gamma_{\ \left \lbrace \beta \sigma \right \rbrace}^{\ \sigma} 
\ \right )
\vspace*{0.3cm} \\
- \ \partial_{\ \alpha} 
\ \left \lbrack
\ e \ Q 
\ \left (
\ g^{\ \varrho \sigma}
\ \gamma_{\ \left \lbrace \varrho \sigma \right \rbrace}^{\ \alpha} 
\ - 
\ g^{\ \alpha \beta}
\ \gamma_{\ \left \lbrace \beta \sigma \right \rbrace}^{\ \sigma} 
\ \right )
\ \right \rbrack
\end{array}
\right \rbrace
\end{array}
\end{equation}

\noindent
We note the form of the second order differential operator
corresponding to $E_{\ \nu \tau}$

\begin{equation}
\begin{array}{l}
\label{eq:18}
E_{\ \nu \tau}^{\ \ \nu' \tau'} \ ( \ g \ ; \ \partial_{\ \varrho} \ ) \ =
\ D_{\ \nu \tau}^{\ \ \nu' \tau'} \ - \ \frac{1}{2} \ g_{\ \nu \tau}
\ D^{\ \nu' \tau'} 
\vspace*{0.3cm} \\
E_{\ \nu \tau}^{\ \ \nu' \tau'} 
\ = 
\ \frac{1}{2}
\ \left (
 \begin{array}{c}
\ \left ( \ \delta_{\ \nu}^{\ \nu'}
\ \delta_{\ \tau}^{\ \tau'}
\ - \ g_{\ \nu \tau} \ g^{\ \nu' \tau'}
\ \right )
\ g^{\ \alpha \beta}
\ \partial_{\ \alpha} \ \partial_{\ \beta} 
\vspace*{0.3cm} \\
\ +
\ g^{\ \nu' \tau'}
\ \partial_{\ \nu} \ \partial_{\ \tau} 
\vspace*{0.3cm} \\
- \ \delta_{\ \nu}^{\ \nu'} 
\ g^{\ \tau' \alpha}
\ \partial_{\ \tau} \ \partial_{\ \alpha} 
- \ \delta_{\ \tau}^{\ \tau'} 
\ g^{\ \nu' \alpha}
\ \partial_{\ \nu} \ \partial_{\ \alpha} 
\vspace*{0.3cm} \\
+ \ g_{\ \nu \tau} \ g^{\ \nu' \alpha} \ g^{\ \tau' \beta}
\ \partial_{\ \alpha} \ \partial_{\ \beta} 
\end{array}
 \right )
\end{array}
\end{equation}

\noindent
The characteristic propagation cone is obtained replacing the partial
derivatives in the expression for $E_{ \nu \tau}^{\ \ \nu' \tau'}$ 
in eq. (\ref{eq:18}) by the directional vector 
$\partial_{ \varrho} \rightarrow \xi_{ \varrho \ }$:

\begin{equation}
\begin{array}{l}
\label{eq:19}
E_{\ \nu \tau}^{\ \ \nu' \tau'} \ ( \ g \ ; \ \xi \ ) \ =
\ \frac{1}{2}
\ \left (
 \begin{array}{c}
\ \left ( \ \delta_{\ \nu}^{\ \nu'}
\ \delta_{\ \tau}^{\ \tau'}
\ - \ g_{\ \nu \tau} \ g^{\ \nu' \tau'}
\ \right )
\ \xi^{\ 2}
\vspace*{0.3cm} \\
\ +
\ g^{\ \nu' \tau'}
\ \xi_{\ \nu} \ \xi_{\ \tau} 
\ + \ g_{\ \nu \tau} \ \xi^{\ \nu'} \ \xi^{\ \tau'}
\vspace*{0.3cm} \\
- \ \delta_{\ \nu}^{\ \nu'} 
\ \xi_{\ \tau}  
\ \xi^{\ \tau'} 
\ - \ \delta_{\ \tau}^{\ \tau'} 
\ \xi_{\ \nu} 
\ \xi^{\ \nu'}
%\vspace*{0.3cm} \\
%- \ g_{\ \nu \tau} \ \xi^{\ \nu'} \ \xi^{\ \tau'}
\end{array}
 \right )
\end{array}
\end{equation}

\noindent
Upon symmetrization of $E_{\ \nu \tau}^{\ \ \nu' \tau'}$ with respect to $^{\ \nu' \tau'}$
we verify the transverse projection condition

\begin{equation}
\begin{array}{l}
\label{eq:20}
\xi^{\ \nu} \ E_{\ \nu \tau}^{\ \ \nu' \tau'} \ =
\ \frac{1}{2}
\ \left (
 \begin{array}{c}
\ \frac{1}{2} 
\left ( 
\ \xi^{\ \nu'}
\ \delta_{\ \tau}^{\ \tau'}
\ + \ \xi^{\ \tau'}
\ \delta_{\ \tau}^{\ \nu'}
\ - \ 2 \ \xi_{\ \tau} \ g^{\ \nu' \tau'}
\ \right )
\ \xi^{\ 2}
\vspace*{0.3cm} \\
\ +
\ g^{\ \nu' \tau'}
\ \xi_{\ \tau} 
\ \xi^{\ 2} 
\ + \ \xi_{\ \tau} \ \xi^{\ \nu'} \ \xi^{\ \tau'}
\vspace*{0.3cm} \\
- \ 
\ \xi_{\ \tau}  
\ \xi^{\ \nu'} 
\ \xi^{\ \tau'} 
\ - \ \frac{1}{2}
\ \left ( 
\ \delta_{\ \tau}^{\ \tau'} 
\ \xi^{\ \nu'}
\ +  
\ \delta_{\ \tau}^{\ \nu'} 
\ \xi^{\ \tau'}
\ \right )
\ \xi^{\ 2} 
%\vspace*{0.3cm} \\
%- \ g_{\ \nu \tau} \ \xi^{\ \nu'} \ \xi^{\ \tau'}
\end{array}
 \right )
\vspace*{0.3cm} \\
\rightarrow \hspace*{0.3cm} 
\xi^{\ \nu} \ E_{\ \nu \tau}^{\ \ \nu' \tau'} \ = \ 0
\end{array}
\end{equation}

\noindent
This allows to illustrate the quantized nature of configuration space variables,
given the quantized nature of the gravitational field, as is evident from the
equations of motion derived from eq. (\ref{eq:12}) ( actio = reactio )

\begin{equation}
\begin{array}{l}
\label{eq:21}
\ \left \lbrack \ D_{\ \mu} \ D_{\ \nu} \ - \ g_{\ \mu \nu} \ D^{\ 2} \ \right \rbrack
\ Q 
\vspace*{0.3cm} \\
+ \ Q \ E_{\ \mu \nu} \ = \ - \ \frac{1}{2} \ \vartheta_{\ \mu \nu}^{\ matter}
\end{array}
\end{equation}

\noindent
The characteristic light cone variables $\xi$ in eqs. (\ref{eq:19}, \ref{eq:20})
appears as tangent vector to the light like geodesics and obey the
geodesic equaion ( eq. \ref{eq:8} )

\begin{equation}
\begin{array}{l}
\label{eq:22}
\dot{\xi}^{\ \nu} \ + \ \gamma_{\ \left \lbrace \mu \varrho \right \rbrace}^{\ \nu} 
\ \xi^{\ \mu} \ \xi^{\ \varrho}
\ = \ 0
\hspace*{0.3cm} ; \hspace*{0.3cm} g_{\ \mu \nu} \ \xi^{\ \mu} \ \xi^{\ \nu} \ = \ 0
%\vspace*{0.3cm} \\
\hspace*{0.3cm} ; \hspace*{0.3cm} 
\dot{X}^{\ \mu} \ = \ \xi^{\ \mu}
\end{array}
\end{equation}

\noindent
While the dot in eq. (\ref{eq:22}) denotes the derivative with respect to a classical
path variable $\tau_{\ cl}$ along the sought light like geodesic, the relations in 
eqs. (\ref{eq:21}, \ref{eq:22}) reveal the configuration space variables
$\xi$ and X as quantized operators, depending together with $\tau_{\ cl}$ 
on an exhaustive set of classical base space variables $y_{\ cl}$.

\noindent
Thus the configuration space variables X, necessarily the arguments of
quantized fields in order to safeguard locality, describe a quantum deformation of the
base space, spanned by $y_{\ cl}$, as is illustrated by target space and
base space in string theory. The associated space geometry is sometimes
referred to as noncommutative \cite{Connes}.

\noindent
These considerations are simple and straightforward, yet the dimensionality
of both target and base space is by no means clearly ten to eleven for the former and
{\it two} for the latter. As a general warning let me state here,
that the noncommutativity of target space variables 
$X^{\ \mu} \ \leftrightarrow \ y_{\ cl}^{\ \nu}$ is encoded in the dynamics, which
influences causally dependent variables and the latter does not allow in general
for any nontrivial algebraic reduction.

\noindent
Looking back at eqs. (\ref{eq:10}) and (\ref{eq:11}) we show the effect
of a Weyl transformation with a {\it constant} scale function f :

\begin{equation}
\begin{array}{l}
\label{eq:23}
e_{\ \mu}^{\ a} 
\ \rightarrow 
\ f \ e_{\ \mu}^{\ a} 
\ = \ \overline{e}_{\ \mu}^{\ a}
\hspace*{0.3cm} ; \hspace*{0.3cm} 
R 
\ \rightarrow 
\ f^{\ -2} \ R
\ = \ \overline{R}
\vspace*{0.3cm} \\
e \ Q \ R 
\ \rightarrow 
\overline{e} \ \overline{Q} \ \overline{R} 
\hspace*{0.3cm} ; \hspace*{0.3cm} 
\overline{Q} \ = \ f^{\ d-2} \ Q 
\end{array}
\end{equation}

\noindent
When the rescaling function f is chosen x-dependent in order to eliminate
the factor $Q \ ( \ \varphi \ )$ in the expression for the action density s
in eq. (\ref{eq:10}) the quadratic dependence of s including the kinetic
terms in ${\cal{L}}_{\ matter}$ is lost.

\noindent
In the flat space limit and $d \ \rightarrow \ 4$ as indicated in eq. (\ref{eq:10})
and neglecting further gravitational effects in the remaining matter
interactions we reduce eq. (\ref{eq:12}) to the (approximate) 
Einstein equations focusing on QCD in flat space-time :

\begin{equation}
\begin{array}{l}
\label{eq:24}
E_{\ \mu \nu} \ = \ - \ 8 \pi \ G_{\ N} \ \vartheta_{\ \mu \nu}
\hspace*{0.3cm} ; \ \mbox{QCD :} \hspace*{0.3cm} 
\left \langle \ \Omega \ \right | 
\ \vartheta_{\ \mu \nu}
\ \left | \ \Omega \ \right \rangle
\ \varepsilon \ g_{\ \mu \nu} \ \approx \ \varepsilon \ \eta_{\ \mu \nu}
\vspace*{0.3cm} \\
g_{\ \mu \nu} \ = \ \eta_{\ \mu \nu} \ + \ h_{\ \mu \nu}
\ \approx \ \eta_{\ \mu \nu}
\hspace*{0.3cm} ; \hspace*{0.3cm} 
\varepsilon \ \rightarrow
\ \varepsilon_{\ QCD}
\vspace*{0.3cm} \\
- \ \varepsilon_{\ QCD}
\ = \ \left ( \ 0.23 \ GeV \ \right )^{\ 4} \ \div
\ \left ( \ 0.28 \ GeV \ \right )^{\ 4} 
\end{array}
\end{equation}

\noindent
In eq. (\ref{eq:23}) $\eta_{\ \mu \nu}$ denotes the conventional Lorentz metric,
normalized to $\eta_{\ 00} \ = \ 1$.

\noindent
The negative vacuum energy density (or positive vacuum pressure) was deduced
first by Shifman, Vainshtain and Zakharov \cite{ShiVaZa} from charmonium
sum rules \cite{Nari} , \cite{SKPM}.
\vspace*{0.1cm}

\noindent
{\bf The dog that did not bark}

\noindent
The content of eq. (\ref{eq:24}) brings about a paradox, since restricting our {\it view}
to $d \ = \ 1 \ + \ 3$ dimensions space time becomes a homogeneous 
space or as the mathematical language says 'inherits a cosmological term' :

\begin{equation}
\begin{array}{l}
\label{eq:25}
R_{\ \mu \nu} \ -
\ \frac{1}{2} \ g_{\ \mu \nu} \ R \ =
\ - \ 8 \ \pi \ G_{\ N} \ \varepsilon \ g_{\ \mu \nu}
\hspace*{0.3cm} \rightarrow \hspace*{0.3cm} 
\vspace*{0.3cm} \\
R_{\ \mu \nu} \ = 
\ \left ( \ 8 \ \pi \ G_{\ N} \ \varepsilon \ / \ ( \ \frac{1}{2} \ d \ - \ 1 \ )
\ \right ) \ g_{\ \mu \nu}
\hspace*{0.3cm} \rightarrow \hspace*{0.3cm} 
\vspace*{0.3cm} \\
R_{\ \left \lbrack \ \alpha \mu \ \right \rbrack 
\ ; \ \left \lbrack \ \beta \ \nu \ \right \rbrack} \ = 
\ \sigma \ a^{\ -2}
\ \left ( 
\ g_{\ \alpha \beta}
\ g_{\ \mu \nu}
\ -
\ g_{\ \alpha \nu}
\ g_{\ \beta \mu}
\ \right )
\hspace*{0.3cm} ; \hspace*{0.3cm} 
\sigma \ = \ \pm \ 1
\vspace*{0.3cm} \\
R_{\ \mu \nu} \ = 
\ ( \ d \ - 1 \ )
\ \sigma \ a^{\ -2}
\ g_{\ \mu \nu}
\hspace*{0.3cm} \rightarrow \hspace*{0.3cm} 
\vspace*{0.3cm} \\
\sigma \ a^{\ -2} \ =
\ \begin{array}{c}
1 
 \vspace*{0.3cm} \\
\hline	\vspace*{-0.3cm} \\
( \ d \ - \ 1 \ ) \ ( \ \frac{1}{2} \ d \ - \ 1 \ )
\end{array}
\ 8 \ \pi \ G_{\ N} \ \varepsilon 
\hspace*{0.3cm} \rightarrow \hspace*{0.3cm} 
\vspace*{0.3cm} \\
\sigma \ = \ \varepsilon \ / \ \left | \ \varepsilon \ \right |
\hspace*{0.3cm} ; \hspace*{0.3cm} 
\sigma_{\ QCD} \ = \ - \ 1
\vspace*{0.3cm} \\
d \ \rightarrow \ 4 
\hspace*{0.3cm} : \hspace*{0.3cm} 
a_{\ QCD} \ = \ \sqrt{\frac{3}{8 \ \pi}}
\ \begin{array}{c}
m_{\ Pl} 
 \vspace*{0.3cm} \\
\hline	\vspace*{-0.3cm} \\
\left | \ \varepsilon_{\ QCD} \ \right |^{\ 1/4}
\end{array}
\ \left | \ \varepsilon_{\ QCD} \ \right |^{\ - 1/4}
\end{array}
\end{equation}

\noindent
The negative sign $\sigma_{\ QCD} \ = \ - \ 1$ selects Anti de Sitter space 

\begin{displaymath}
AdS \ 4 \ = \ SO \ ( \ 2 , 3 \ ) \ / \ SO \ ( \ 1 , 3 \ ) 
\end{displaymath}

\noindent
as homogeneous four dimensional space time. 

\noindent
The positive sign would select de Sitter space

\begin{displaymath}
dS \ 4 \ = \ SO \ ( \ 1 , 4 \ ) \ / \ SO \ ( \ 4 \ ) 
\end{displaymath}

\noindent
The curvature radius becomes

\begin{equation}
\begin{array}{l}
\label{eq:26}
a_{\ QCD} \ = 
\ 13.5 \ km 
\ \left (
\ \begin{array}{c}
0.25 \ GeV
 \vspace*{0.3cm} \\
\hline	\vspace*{-0.3cm} \\
\left | \ \varepsilon_{\ QCD} \ \right |^{\ 1/4}
\end{array}
\ \right )^{\ 2}
\end{array}
\end{equation}

\noindent
{\it New York would not be the same.}

\noindent
There is no doubt that the above curvature is not present in the (almost) flat
four dimensions, and an interesting paradox arises. Yet if the above deduced
curvature in the form

\begin{equation}
\begin{array}{l}
\label{eq:27}
\left ( \ \sigma \ a^{\ -2} \ \right )_{\ QCD}
\end{array}
\end{equation}

\noindent
does not curve the four known dimensions it should then curve some other space.

\noindent
{\bf cosmological acceleration and negative pressure}

\noindent
If we interpret the large scale cosmological
{\it acceleration} found by the large z supernova Ia redshift surveys
\cite{Riess} , \cite{Perlm} as due to a cosmological constant (constant in time) 
then it follows 

\begin{equation}
\begin{array}{l}
\label{eq:28}
\sigma_{\ cos} \ = \ + \ 1
\hspace*{0.3cm} \leftrightarrow \hspace*{0.3cm} 
\sigma_{\ QCD} \ = \ - \ 1
 \vspace*{0.3cm} \\
a_{\ cos} \ = 
\ \begin{array}{c}
c
 \vspace*{0.3cm} \\
\hline	\vspace*{-0.3cm} \\
\sqrt{\Omega_{\ cos}} \ H_{\ 0}
\end{array}
\ = \ 2.2 \ 10^{\ 23} \ km
\left .
\ \left ( \ \begin{array}{c}
0.7
 \vspace*{0.3cm} \\
\hline	\vspace*{-0.3cm} \\
\Omega_{\ cos} 
\end{array}
\ \right )^{\ 1/2}
\ \right / \ h_{\ 50}
 \vspace*{0.3cm} \\
h_{\ 50} \ = \ H_{\ 0} \ / \ \left ( \ 50 \ km / sec / Mpsec \ \right )
\end{array}
\end{equation}

\noindent
In eq. (\ref{eq:28}) $H_{\ 0}$ denotes the Hubble expansion parameter at
present and 
$\Omega_{\ cos} \ = \ \varepsilon \ / \ \varepsilon_{\ cr}$ where the
critical energy density is 

\begin{equation}
\begin{array}{l}
\label{eq:29}
\varepsilon_{\ cr} \ = 
\ 3 \ ( \ H_{\ 0} \ )^{\ 2}\ / \ ( \ 8 \ \pi \ G_{\ N} \ )
\end{array}
\end{equation}

\noindent
The large curvature radius $a_{\ cos}$ as deduced in eq. (\ref{eq:28})
while acceptable phenomenologically together with the notable change
of sign $\sigma_{\ cos} \ = \ + \ 1$ does not seem to resolve the
paradoxical situation with respect to the cosmological term,
which we summarize in table 1 :

\clearpage

\begin{table}[ht]
\[
\begin{array}{c@{\hspace*{0.6cm}}c@{\hspace*{0.4cm}}
c@{\hspace*{0.4cm}}c}\\ \hline
\mbox{origin} & \sigma & \sim \ |\mbox{energy density}|^{\ 1/4} 
& \sim \ \mbox{curvature radius} (\mbox{km})
\\ \hline
\mbox{QCD} & - 1 & 0.25 \ GeV & 13.5 
\vspace*{0.1cm}   \\
\mbox{cosmos} & + 1 & 2 \ 10^{\ -3} \ eV & 2.2 \ 10^{\ 23}
\vspace*{0.1cm} \\
\hline
\end{array}
\]
\caption{Survey of candidate cosmological curvatures.}
\label{tabcos}
\end{table}

\section*{2 + 3 QED and QCD}

\noindent
We restrict this section to the structure of the QCD Lagrangean and
the ensuing two central anomaliesr, inasfar as these are relevant for the
study of vacuum energy density and potential induced cosmological terms.

\noindent
For a review see e.g. refs. \cite{Dok}, \cite{leut}, \cite{PMCz}.
The Lagrangean (flat space) density with Fermi gauge parameter $\eta$ is of the form

\begin{equation}
\begin{array}{l}
\label{eq:30}
{\cal{L}}_{\ QCD} \ =
\ \left \lbrace
\ \begin{array}{l}
- \ 
\begin{array}{c}
1
 \vspace*{0.3cm} \\
\hline	\vspace*{-0.4cm} \\
2 \ \eta \ g^{\ 2} 
\end{array}
\ \ C^{\ A} \ C^{\ A}
\ + \ 2 \ tr \ \partial^{\ \mu} \ \overline{c} \ D_{\ \mu} \ c 
\vspace*{0.3cm} \\
- \ 
\begin{array}{c}
1
 \vspace*{0.3cm} \\
\hline	\vspace*{-0.4cm} \\
4 \ g^{\ 2} 
\end{array}
\ F_{\ \mu \nu}^{\ A} \ F^{\ \mu \nu \ A}
\ + \ \overline{q} \ i \ \gamma^{\ \mu} \ D_{\ \mu} \ q
%\vspace*{0.3cm} \\
\ - \ m_{\ q} \ \overline{q} \ q
\end{array}
\ \right .
\vspace*{0.3cm} \\
D_{\ \mu} \ = \ \partial_{\ \mu} \ + \ {\cal{W}}_{\ \mu}
\hspace*{0.3cm} ; \hspace*{0.3cm} 
{\cal{W}}_{\ \mu} \ = \ i \ {\cal{D}}^{\ A} \ V_{\ \mu}^{\ A}
\vspace*{0.3cm} \\
C^{\ A} \ = \ \partial_{\ \mu} \ V_{\ \mu}^{\ A}
\hspace*{0.3cm} ; \hspace*{0.3cm} 
c \ = \ i \ {\cal{D}}^{\ A} c^{\ A}
\hspace*{0.3cm} ; \hspace*{0.3cm} 
\overline{c} \ = \ -i \ {\cal{D}}^{\ A} c^{\ * \ A}
\vspace*{0.3cm} \\
F_{\ \mu \nu}^{\ A} \ = 
\ \partial_{\ \nu} \ V_{\ \mu}^{\ A}
\ - \ \partial_{\ \mu} \ V_{\ \nu}^{\ A}
\ - \ f_{\ A B C} \ V_{\ \nu}^{\ B}  V_{\ \mu}^{\ C} 
\end{array}
\end{equation}

\noindent
In eq. (\ref{eq:30}) ${\cal{D}}_{\ A}$ denotes the fundamental representation of 
$SUN_{\ c}$ with $N_{\ c} \ \rightarrow \ 3$ and the conventional normalization

\begin{equation}
\begin{array}{l}
\label{eq:31}
tr \ {\cal{D}}^{\ A} \ {\cal{D}}^{\ B} \ = \frac{1}{2} \ \delta_{\ A B}
\hspace*{0.3cm} ; \hspace*{0.3cm} 
\left \lbrack \ {\cal{D}}^{\ A} \ , \ {\cal{D}}^{\ B} \ \right \rbrack
\ = \ i \ f_{\ A B C} \ {\cal{D}}^{\ C}
\vspace*{0.3cm} \\
\mbox{with} \ D_{\ \mu} \ c \ = \partial_{\ \mu} \ c \ + 
\left \lbrack \ {\cal{W}}_{\ \mu} \ , \ c \ \right \rbrack
\end{array}
\end{equation}

\noindent
The central anomalies become 'trace' or conformal and singlet axial current anomalies 
in flat d = 4 space time

\begin{equation}
\begin{array}{l}
\label{eq:32}
\vartheta^{\ \mu}_{\ \mu} \ = 
 \begin{array}{c}
b_{\ 1} 
 \vspace*{0.3cm} \\
\hline	\vspace*{-0.4cm} \\
8 \ \pi^{\ 2} 
\end{array}
\hspace*{0.2cm}
\left ( \ - \ \frac{1}{4} \ 
\ F_{\ \mu \nu}^{\ A} \ F^{\ \mu \nu \ A}
\ \right )_{\ ren.gr.inv.}
\ + \ m_{\ q} \ \overline{q} \ q
\vspace*{0.3cm} \\
j_{\ \mu}^{\ 5} \ = \ \sum_{\ q} \ \overline{q} 
\ \gamma_{\ \mu} 
\ \gamma_{\ 5}^{\ R} \ q 
\hspace*{0.3cm} ; \hspace*{0.3cm} 
\gamma_{\ 5}^{\ R} \ = \ - i \ \gamma_{\ 0} 
\ \gamma_{\ 1} \ \gamma_{\ 2} \ \gamma_{\ 3} 
\ \rightarrow \ \gamma_{\ 5}
\vspace*{0.3cm} \\
\partial^{\ \mu} \ j_{\ \mu}^{\ 5} \ =
\ 2 \ n_{\ fl}
\ \begin{array}{c}
1 
 \vspace*{0.3cm} \\
\hline	\vspace*{-0.4cm} \\
8 \ \pi^{\ 2} 
\end{array}
\hspace*{0.2cm}
\left ( \ \frac{1}{4} \ 
\ F_{\ \mu \nu}^{\ A} \ \widetilde{F}^{\ \mu \nu \ A}
\ \right )_{\ ren.gr.inv.}
\ + \ 2 \ m_{\ q}
\ \overline{q} \ i \ \gamma_{\ 5} \ q
 \vspace*{0.3cm} \\
\left ( \ \frac{1}{4} \ 
\ F_{\ \mu \nu}^{\ A} \ \widetilde{F}^{\ \mu \nu \ A}
\ \right )_{\ ren.gr.inv.}
\ \equiv \ ch_{\ 2} \ ( \ F \ )
\hspace*{0.3cm} 8 \ \pi^{\ 2}
\end{array}
\end{equation}

\noindent
In eq. (\ref{eq:32}) $n_{\ fl}$ denotes the number of quark flavors
and $b_{\ 1}$ refers to the (perturbative) rescaling function

\begin{equation}
\begin{array}{l}
\label{eq:33}
b \ ( \ \kappa \ ) \ = \ - \ \beta \ ( \ g \ ) \ / \ g \ =
\ \sum_{\ 1}^{\ \infty} \ b_{\ n} \ \kappa^{\ n}
\hspace*{0.3cm} ; \hspace*{0.3cm} 
\kappa \ =
\ \begin{array}{c}
1 
 \vspace*{0.3cm} \\
\hline	\vspace*{-0.4cm} \\
16 \ \pi^{\ 2} 
\end{array}
\hspace*{0.2cm}
g^{\ 2}
 \vspace*{0.3cm} \\
 b_{\ 1} \ = \ \frac{11}{3} \ C_{\ 2} \ ( \ G \ ) \ - \ \frac{2}{3} \ n_{\ fl}
 \vspace*{0.3cm} \\
%\hspace*{0.3cm} ; \hspace*{0.3cm} 
 b_{\ 2} \ = \ \frac{34}{3} 
 \ \left ( \ C_{\ 2} \ ( \ G \ ) \ \right )^{\ 2}
 - \ 2 \ C_{\ 2} \ ( \ q \ ) \ n_{\ fl}
 - \ \frac{10}{3} \ C_{\ 2} \ ( \ G \ ) \ n_{\ fl}
 \vspace*{0.3cm} \\
 \cdots
\end{array}
\end{equation}

\noindent
In eq. (\ref{eq:33}) $C_{\ 2} \ ( \ G \ )$, $C_{\ 2} \ ( \ q \ )$
denote the second Casimir invariant for the adjoint ( G ) and the
defining ( q ) representations of the gauge group respectively.

\noindent
For $G \ = \ SUN_{\ c}$ and $N_{\ c} \ \rightarrow \ 3$ these quantities become

\begin{equation}
\begin{array}{l}
\label{eq:34}
C_{\ 2} \ ( \ G \ ) \ \rightarrow \ N_{\ c} \ \rightarrow \ 3
\hspace*{0.3cm} ; \hspace*{0.3cm} 
C_{\ 2} \ ( \ q \ ) \ \rightarrow 
\ \begin{array}{c}
N_{\ c}^{\ 2} \ - \ 1 
 \vspace*{0.3cm} \\
\hline	\vspace*{-0.4cm} \\
2 \ \N_{\ c} 
\end{array}
\ \rightarrow \ \frac{4}{3}
 \vspace*{0.3cm} \\
 b_{\ 1} \ \rightarrow \ 11 \ - \ \frac{2}{3} \ n_{\ fl}
\hspace*{0.3cm} , \hspace*{0.3cm} 
 b_{\ 2} \ \rightarrow \ 102 \ - \ \frac{38}{3} \ n_{\ fl}
\end{array}
\end{equation}

\noindent
The 'trace' $\vartheta^{\ \mu}_{\ \mu}$ in eq. (\ref{eq:32})
relates to the 'base' condensates and to the vacuum energy density 
$\varepsilon_{\ QCD} \ = \ - \ p_{\ 0}$ defined in eq. (\ref{eq:24})

\begin{equation}
\begin{array}{l}
\label{eq:35}
\left ( \ \frac{1}{4} \ 
\ F_{\ \mu \nu}^{\ A} \ F^{\ \mu \nu \ A}
\ \right )_{\ ren.gr.inv.}
\ \rightarrow 
\ \frac{1}{4} \ 
\ F_{\ \mu \nu}^{\ A} \ F^{\ \mu \nu \ A}
 \vspace*{0.3cm} \\
 \left \langle \ \Omega \ \right |
\ \frac{1}{4} \ 
\ F_{\ \mu \nu}^{\ A} \ F^{\ \mu \nu \ A}
\ \left | \ \Omega \ \right \rangle
\ = \ {\cal{B}}^{\ 2}
\ = \ \left \lbrace
\ \begin{array}{l}
0.250 \ GeV^{\ 4} \ \mbox{\cite{Nari}}
 \vspace*{0.3cm} \\
0.125 \ GeV^{\ 4} \ \mbox{\cite{ShiVaZa}}
\end{array}
\right .
 \vspace*{0.3cm} \\
 \left \langle \ \Omega \ \right |
 \ \left ( \ - \ \overline{u} \ u \ \right )
\ \left | \ \Omega \ \right \rangle
\ = \ f_{\ \pi}^{\ 2} \ M_{\ 0}
\ \simeq \ \left ( \ 0.24 \ GeV \ \right )^{\ 3}
 \vspace*{0.3cm} \\
 M_{\ 0} \ =
\ \begin{array}{c}
m_{\ \pi}^{\ 2} 
 \vspace*{0.3cm} \\
\hline	\vspace*{-0.4cm} \\
m_{\ u} \ + \ m_{\ d}
\end{array}
\hspace*{0.2cm}
\ \simeq \ 1.3 \ GeV
\end{array}
\end{equation}

\noindent
The quantity $M_{\ 0}$ in eq. (\ref{eq:35}) is 'scale' dependent and corresponds
to that scale here where $m_{\ u} \ + \ m_{\ d} \ = \ 14 \ MeV$.

\noindent
Then the spontaneous QCD parameters $\varepsilon_{\ QCD} \ = \ - \ p_{\ 0}$ 
( eqs. \ref{eq:24} , \ref{eq:35} ) become restricting QCD to the three
light flavors u, d, s

\begin{equation}
\begin{array}{l}
\label{eq:36}
 \left \langle \ \Omega \ \right |
 \ \vartheta_{\ \mu \nu}^{\ QCD}
\ \left | \ \Omega \ \right \rangle
\ = \ \varepsilon_{\ QCD} \ \eta_{\ \mu \nu}
\hspace*{0.3cm} ; \hspace*{0.3cm} 
- \ \varepsilon_{\ QCD} \ = \ p_{\ 0} 
 \vspace*{0.3cm} \\
 p_{\ 0} \ =
\ \begin{array}{c}
9 
 \vspace*{0.3cm} \\
\hline	\vspace*{-0.4cm} \\
32 \ \pi^{\ 2}
\end{array}
\hspace*{0.2cm}
 {\cal{B}}^{\ 2}
 \ + \ \sim \ f_{\ \pi}^{\ 2}
 \ \left ( \ \frac{1}{2} \ m_{\ \pi}^{\ 2} \ + \ m_{\ K}^{\ 2} \ \right )
 \ = \ \left ( 0.23 \ \div \ 0.28 \ GeV \right )^{ 4}
\end{array}
\end{equation}

\noindent
The range of values for the vacuum pressure $p_{\ 0}$ in eq. (\ref{eq:36})
corresponds to the lower and higher value for the spontaneous parameter
${\cal{B}}^{\ 2}$ given in eq. (\ref{eq:35}).

\begin{figure}[ht]
\begin{center}
\mbox{\epsfig{file=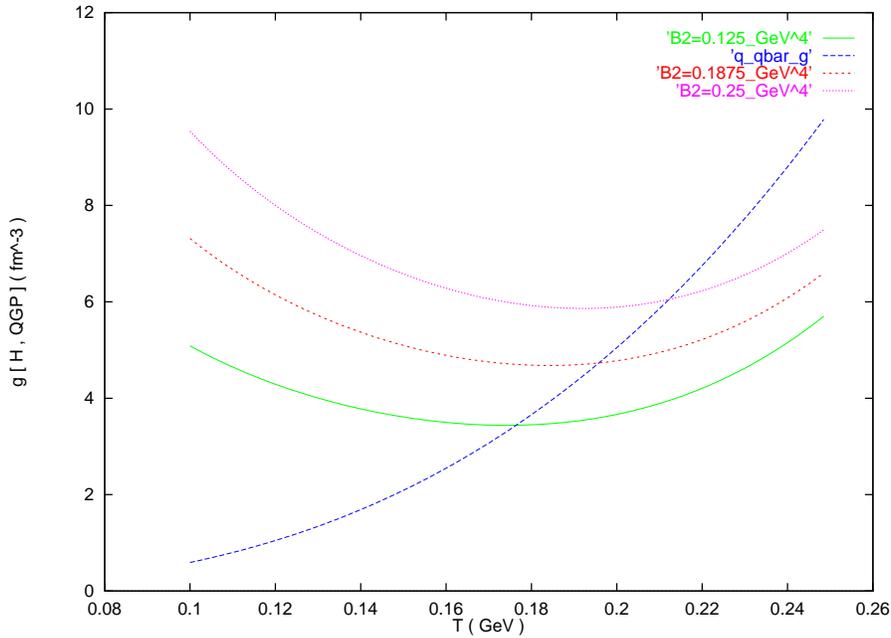,width=120mm}}
\end{center}
\caption{
Gibbs potentials as a function of the temperature for three values
of the gluon condensate in the ground state,
and for free quark flavors u, d, s and gluons.
}
\label{gbfig3}
\end{figure}

\noindent
This led to the attempt to map out the thermodynamically the phase structure
of QCD in nucleus-nucleus, $p-\overline{p}$ and $e^{+}-e^{-}$ collisions
in ref. \cite{SKPM}, extrapolating all systems to zero chemical
potentials. The Gibbs potentials $g \ ( \ T \ ) \ = \ p \ ( \ T \ ) \ / T$
for the hadronic ($g^{\ H}$) and the plasma ($g^{\ QGP}$) phase,
where hadronic interactions in the hadron phase are replaced by a representative
set of noninteracting resonances, while in the plasma phase u, d , s flavored quarks
are endowed with masses and further interactions are neglectedm
are shown in figure \ref{gbfig3}.

\begin{figure}[ht]
\begin{center}
\mbox{\epsfig{file=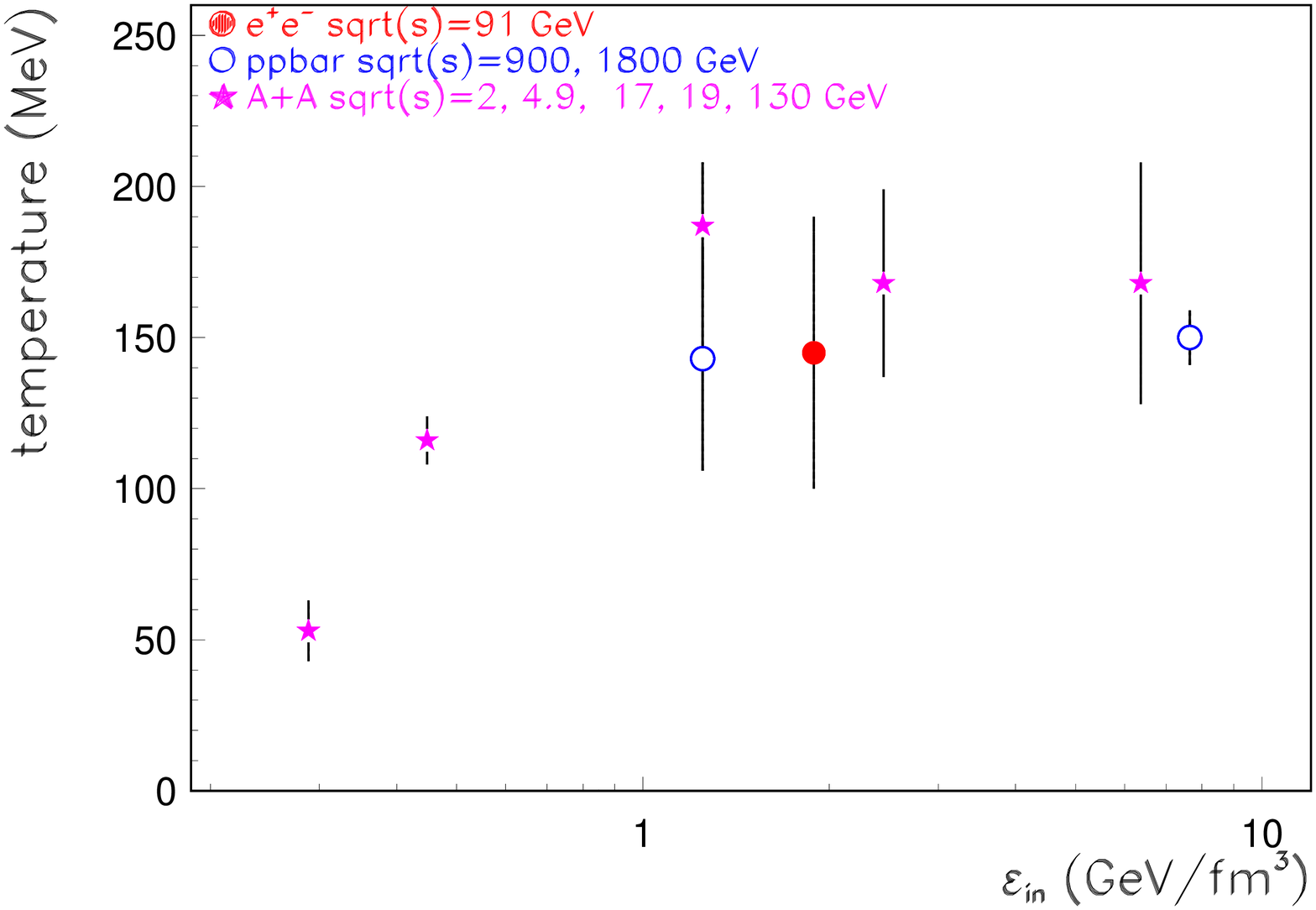,width=120mm}}
\end{center}
\caption{
Temperature at  chemical freeze-out and for zero fugacities as a function of
the initial energy density for several nucleus+nucleus, hadron+hadron and
lepton+lepton collisions.
We demand for the fits confidence level $>$ 10\%.
}
\label{t3}
\end{figure}

\noindent
The determination of the temperature characterizing various collision systems are
shown in figure \ref{t3}.

\noindent
Equating the pressure for the two phases yields for the critical temperature
at zero chemical potentials the estimate

\begin{equation}
\begin{array}{l}
\label{eq:37}
T_{\ cr} \ = \ 194 \ \pm \ 18 MeV
\end{array}
\end{equation}

\noindent
In the error for $T_{\ cr}$ in eq. (\ref{eq:37}) the systematic errors
both in theoretical approximations and in the thermal description of the scattering
dynamics are included.

\section*{4 Completing the standard model}
\vspace*{0.1cm}

\noindent
In this section I concentrate on spontaneous gauge symmetry breaking \cite{Higgs},
\cite{EngBr}, \cite{JEDR}.

\noindent
The new feature is the appearance of a multiplet of scalar elementary fields.
The minimal single doublet represents a quaternion, two doublets an octonion:

\begin{equation}
\begin{array}{l}
\label{eq:38}
\left (
\ \begin{array}{c}
\nu_{\ e}
\vspace*{0.3cm} \\
e^{\ -}
\end{array}
\ \right )
\ \rightarrow 
\ \left (
\ \begin{array}{c}
\varphi^{\ 0}
\vspace*{0.3cm} \\
\varphi^{\ -}
\end{array}
\ \right )
\ \rightarrow 
\ \Phi \ =
\ \left (
\ \begin{array}{cr}
\varphi^{\ 0} & - \ \left ( \ \varphi^{\ -} \ \right )^{\ *} 
\vspace*{0.3cm} \\
\varphi^{\ -} & \left ( \ \varphi^{\ 0} \ \right )^{\ *} 
\end{array}
\ \right )
\vspace*{0.3cm} \\
\sqrt{2} \ \Phi \ =
\ \left (
\ \begin{array}{lr}
\Phi_{\ 0} \ - \ i \ \Phi_{\ 3} & - \ \Phi_{\ 2} \ - \ i \ \Phi_{\ 1}  
\vspace*{0.3cm} \\
\Phi_{\ 2} \ - \ i \ \Phi_{\ 1} & \ \Phi_{\ 0} \ + \ i \ \Phi_{\ 3}  
\end{array}
\ \right )
\ = \ \Phi_{\ 0} \ + \ \frac{1}{i} \ \vec{\sigma} \ \vec{\Phi}
\ = \ \xi
\end{array}
\end{equation}

\noindent
The invariant of the electroweak gauge group $SU2_{\ L} \ U1_{\ {\cal{Y}}}$

\begin{equation}
\begin{array}{l}
\label{eq:39}
\overline{\xi} \ \xi \ = \ 2 \ \varphi^{\ \dagger} \ \varphi
\ = \ \Phi_{\ 0}^{\ 2} \ + \ \vec{\Phi}^{\ 2} \ = \Phi^{\ 2}
\end{array}
\end{equation}

\noindent
exhibits the enlarged $SU2_{\ L} \ \times \ SU2'$ invariance.

\noindent
The scalar self interaction generates after $\varepsilon_{\ QCD}$ the
next higher spontaneous effect in mass scale. The former is responsible for
most of the nucleon mass ( $\sim \ 900$ out of $938$ MeV ) , while 
the latter generates mass for all quark and charged lepton flavors, the W and Z gauge
bosons and minimally the one remaining scalar Higgs boson :

\begin{equation}
\begin{array}{l}
\label{eq:40}
V \ ( \ \Phi \ ) \ = \ - \ \frac{1}{2} \ \mu^{\ 2} \ \Phi^{\ 2}
\ + \ \frac{1}{8} \ \lambda \ \left ( \ \Phi^{\ 2} \ \right )^{\ 2} \ ( \ + \ constant \ )
\vspace*{0.3cm} \\
\Phi_{\ 0} \ = \ v \ + \ H \ ( \ x \ )
%\hspace*{0.3cm} ; \hspace*{0.3cm} 
\vspace*{0.3cm} \\
v \ = \ \sqrt{2} \ \mu \ / \ \sqrt{\lambda} \ = 
\ \left ( \ \sqrt{2} \ G_{\ F} \ \right )^{\ -1/2} \ \simeq \ 246.2 \ GeV
\vspace*{0.3cm} \\
V \ ( \ v \ + \ H \ ) \ = \ - \ \frac{1}{2} \ \mu^{\ 2} \ v^{\ 2}
\ + \ \frac{1}{8} \ \lambda \ v^{\ 4} \ + \ \frac{1}{2}
\ \left ( \ - \mu^{\ 2} \ + \ \frac{3}{2} \ \lambda \ v^{\ 2} \ \right ) \ H^{\ 2}
\vspace*{0.3cm} \\
\rightarrow \ m_{\ H}^{\ 2} \ = \ 2 \ \mu^{\ 2} \ = \ \lambda \ v^{\ 2}
%\hspace*{0.3cm} ; \hspace*{0.3cm} 
\vspace*{0.3cm} \\
V \ ( \ v \ ) \ = 
\ \begin{array}[t]{l}
 - \ \frac{1}{8} \ m_{\ H}^{\ 2} \ v^{\ 2} \ ( \ + \ constant \ )
\vspace*{0.3cm} \\
\ \rightarrow \ - \ \varepsilon^{\ \Phi}_{\ vac} \ = \  \left ( \ 104.3 \ GeV \ \right )^{\ 4} 
\ \left ( \ m_{\ H} \ / \ 125 \ GeV \ \right )^{\ 2}
\end{array}
\end{array}
\end{equation}

\noindent
Upon neglecting the free constant in the definition of the $\Phi$ potential
we obtain a negative induced curvature as in the case of QCD with
a correspondingly smaller curvature radius

\begin{equation}
\begin{array}{l}
\label{eq:41}
\left (
\ \begin{array}{c}
\varepsilon^{\ \Phi}_{\ vac}
 \vspace*{0.3cm} \\
\hline	\vspace*{-0.4cm} \\
\varepsilon_{\ QCD}
\end{array}
\ \right )^{\ 1/2} \ \sim \ ( \ 417 \ )^{\ 2}
\ \left ( \ m_{\ H} \ / \ 125 \ GeV \ \right )^{\ 2}
\vspace*{0.3cm} \\
a_{\ \Phi \ 4-d} \ = \ 417^{\ -2} \ a_{\ QCD} \ \sim \ \frac{1}{2} \ cm
\hspace*{0.3cm} \mbox{for} \hspace*{0.3cm} 
m_{\ H} \ = \ 125 \ GeV
\end{array}
\end{equation}

\noindent
" $\cdots$ they prove the impossible and disprove the obvious $\cdots$ ".

\noindent
Table \ref{tabcos} thus obtains a new entry :

\begin{table}[ht]
\[
\begin{array}{c@{\hspace*{0.6cm}}c@{\hspace*{0.4cm}}
c@{\hspace*{0.4cm}}c}\\ \hline
\mbox{origin} & \sigma & \sim \ |\mbox{energy density}|^{\ 1/4} 
& \sim \ \mbox{curvature radius} (\mbox{km})
\\ \hline
\mbox{QCD} & - 1 & 0.25 \ GeV & 13.5 
\vspace*{0.1cm}   \\
\mbox{cosmos} & + 1 & 2 \ 10^{\ -3} \ eV & 2.2 \ 10^{\ 23}
\vspace*{0.1cm} \\
\mbox{el. weak} & - 1 & 105 \ GeV & 0.5 \ 10^{\ -5}
\vspace*{0.1cm} \\
\hline
\end{array}
\]
\caption{Survey of candidate cosmological curvatures.}
\label{tabcosw}
\end{table}

\section*{5 + 6 Unification of charges and susy}
\vspace*{0.1cm}

\noindent
{\bf Susy and vacuum energy density}
\vspace*{0.1cm}

\noindent
As in the preceding section we focus on the aspect of spontaneous vacuum energy
density in the context of initially flat 4-d rigid supersymmetry.

\noindent
For a recent overview see e.g. \cite{JEllis}.
\vspace*{0.1cm}

\noindent
We consider the 'once local' form of the algebra of N susy charges
\cite{LBPM}

\begin{equation}
\begin{array}{l}
\label{eq:42}
\left \lbrace 
\ Q^{\ i}_{\ \alpha}
\ , \ Q^{\ * k}_{\ \dot{\beta}}
\ \right \rbrace
\ = \ \delta^{\ i k} \ \sigma^{\ \nu}_{\ \alpha \dot{\beta}} \ P_{\ \nu} 
\hspace*{0.3cm} ; \hspace*{0.3cm} 
i,k \ = \ 1 \cdots N
\ \rightarrow
\vspace*{0.3cm} \\
%\hspace*{0.3cm} ; \hspace*{0.3cm} 
\left \lbrace 
\ j^{\ i}_{\ \mu \alpha} \ ( \ x \ )
\ , \ Q^{\ * k}_{\ \dot{\beta}}
\ \right \rbrace
\ = \ \delta^{\ i k} \ \sigma^{\ \nu}_{\ \alpha \dot{\beta}} 
\ \vartheta_{\ \mu \nu} \ ( \ x \ )
\end{array}
\end{equation}

\noindent
In eq. (\ref{eq:42}) $j^{\ i}_{\ \mu \alpha} \ ( \ x \ )$ denote the N local
and conserved spinorial susy currents, whereas 
$\vartheta_{\ \mu \nu} \ ( \ x \ )$ is the (conserved) energy momentum density operator,
in d = 1 + 3 flat dimensions.

\noindent
An {\it eventual} spontaneous vacuum energy density 
of the form defined in eq. (\ref{eq:24})

\begin{equation}
\begin{array}{l}
\label{eq:43}
\mbox{susy :} \hspace*{0.3cm} 
\left \langle \ \Omega \ \right | 
\ \vartheta_{\ \mu \nu}
\ \left | \ \Omega \ \right \rangle
\ = \ \varepsilon_{\ susy} \ g_{\ \mu \nu} \ \approx \ \varepsilon_{\ susy} \ \eta_{\ \mu \nu}
\end{array}
\end{equation}

\noindent
in the susy environment leads not only to the spontaneous breakdown of all
N supersymmetries but to a most notable change of sign, relative to the
case encountered in QCD and in the Higgs effect

\begin{equation}
\begin{array}{l}
\label{eq:44}
\varepsilon_{\ susy} \ \ge \ 0 
\ \leftrightarrow
\vspace*{0.3cm} \\
\left \langle \ \Omega \ \right | 
\ \left \lbrace 
\ j^{\ i}_{\ \mu \alpha} \ ( \ x \ )
\ , 
\ j^{\ * k}_{\ \nu \dot{\beta}} \ ( \ y \ )
\ \right \rbrace
\ \left | \ \Omega \ \right \rangle
\ = \ \delta^{\ i k} \ \Gamma_{\ \mu \nu \varrho} \ ( \ z \ )
\ \sigma^{\ \varrho}_{\ \alpha \dot{\beta}} 
\hspace*{0.3cm} ; \hspace*{0.1cm} 
z  =  x  -  y
\vspace*{0.3cm} \\
\Gamma_{\ \mu \nu \varrho} \ ( \ z \ )
\ = \ \left ( \ 2 \pi \ \right )^{\ -3} 
\ {\displaystyle{\int}} \ d^{\ 4} \ q 
\ \varepsilon \ ( \ q^{\ 0} \ )
\ \widetilde{\Gamma}_{\ \mu \nu \varrho} \ ( \ q \ , \ \varepsilon_{\ susy} \ )
\vspace*{0.3cm} \\
\widetilde{\Gamma}_{\ \mu \nu \varrho} \ ( \ q \ , \ \varepsilon_{\ susy} \ )
\ =
\ \varepsilon_{\ susy} \ \delta \ ( \ q^{\ 2} \ )
\ \gamma_{\ \mu \nu \varrho} \ ( \ q \ )
\ + \ \Delta \ \widetilde{\Gamma}_{\ \mu \nu \varrho} \ ( \ q \ , \ \varepsilon_{\ susy} \ )
\vspace*{0.3cm} \\
\gamma_{\ \mu \nu \varrho} \ ( \ q \ )
\ =
\ \eta_{\ \mu \varrho} \ q_{\ \nu}
\ + \ \eta_{\ \nu \varrho} \ q_{\ \mu}
\ - \ \eta_{\ \mu \nu} \ q_{\ \varrho}
\end{array}
\end{equation}

\noindent
In the second last relation in eq. (\ref{eq:44})
the leading spectral singularity \\
( $\propto \ \delta \ ( \ q^{\ 2} \ )$ )
arises - - for $\varepsilon_{\ susy} \ > \ 0$ - - 
from N goldstino modes, as a consequence of the (universal) spontaneous breaking of 
all N supersymmetries. $\Delta \ \widetilde{\Gamma}$ denotes all residual contribution
to the susy current current correlation function, less singular for 
$q^{\ 2} \ \rightarrow \ 0$.

\noindent
The condition $\varepsilon_{\ susy} \ \ge \ 0$ follows from the K\"{a}llen-Lehmann
representation for a local anticommutator and the positivity of 
$j \ j^{\ *}$ and $ j^{\ *} \ j $ products.

\noindent
We note the Christoffel-symbol like structure of $\gamma_{\ \mu \nu \varrho}$
as displayed in eq. (\ref{eq:44}), which is not accidental.
\vspace*{0.1cm}

\noindent
With respect to primordial cosmological inflation {\it and} the apparent
present day remnant as shown in table \ref{tabcosw} the sign
of $\varepsilon_{\ susy}$ is the same.
\vspace*{0.2cm}

\noindent
{\bf Unification of charges $SU3_{\ c} \times SU2_{\ L} \times U1_{\ {\cal{Y}}}$}
\vspace*{0.1cm}

\noindent
The three running coupling constants
$\overline{g}_{\ k} \ ( \ \mu^{\ 2} \ ) \ ; \ k \ = \ 1,2,3$

\begin{displaymath}
\begin{array}{l}
\overline{g}_{\ 3} \ \leftrightarrow \ SU3_{\ c}
\hspace*{0.2cm} ; \hspace*{0.2cm} 
%\vspace*{0.2cm} \\
\overline{g}_{\ 2} \ \leftrightarrow \ SU2_{\ L}
\hspace*{0.2cm} ; \hspace*{0.2cm} 
%\vspace*{0.2cm} \\
\overline{g}_{\ 1} \ \leftrightarrow \ U1_{\ {\cal{Y}}}
\end{array}
\end{displaymath}

\noindent
defined in the $\overline{MS}$ scheme
and normalized to a common Casimir operator of one chiral fermion family
are defined as 

\begin{equation}
\begin{array}{l}
\label{eq:45}
\overline{\alpha}_{\ 1} \ = \ \frac{5}{3} \ \tan^{\ 2} \ \overline{\vartheta}_{\ W}
\ \overline{\alpha}_{\ 2}
\hspace*{0.2cm} ; \hspace*{0.2cm} 
%\vspace*{0.3cm} \\
\overline{\alpha}_{\ 2 (3)} \ = \ \overline{g}_{\ 2 (3)}^{\ 2} \ / \ ( \ 4 \pi \ )
\vspace*{0.3cm} \\
\overline{\alpha}_{\ 1} \ = \ \frac{5}{3} 
\ \left ( \ 1 \ / \ \cos^{\ 2} \ \overline{\vartheta}_{\ W} \ \right )
\ \overline{\alpha}_{\ em}
\hspace*{0.2cm} ; \hspace*{0.2cm} 
\overline{\alpha}_{\ 3} \ = 
\ \left ( \ 1 \ / \ \sin^{\ 2} \ \overline{\vartheta}_{\ W} \ \right )
\ \overline{\alpha}_{\ em}
\end{array}
\end{equation}

\noindent
We thus pick up the three coupling constants at the scale $\mu \ = \ m_{\ Z}$

\begin{equation}
\begin{array}{l}
\label{eq:46}
\mbox{at} \ m_{\ Z} \ : \ \sin^{\ 2} \ \overline{\vartheta}_{\ W} \ = 
\ 0.231078 \ \footnote{This value was derived from precision data analysis
by W. Marciano \cite{WM}}
\hspace*{0.2cm} ; \hspace*{0.2cm} 
\left ( \ \overline{\alpha}_{\ em} \ \right )^{\ -1} 
\ = \ 127.934(27)
\vspace*{0.3cm} \\
\left ( \ \overline{\alpha}_{\ 1} \ \right )^{\ -1} 
\ = \ 59.02 \ (2)
\hspace*{0.2cm} ; \hspace*{0.2cm} 
\left ( \ \overline{\alpha}_{\ 2} \ \right )^{\ -1} 
\ = \ 29.56 \ (1)
\hspace*{0.2cm} ; \hspace*{0.2cm} 
\left ( \ \overline{\alpha}_{\ 3} \ \right )^{\ -1} 
\ = \ 8.47 \ (14)
\end{array}
\end{equation}

\noindent
The rescaling equations are

\begin{equation}
\begin{array}{l}
\label{eq:47}
\overline{\kappa}_{\ k} \ = \ \frac{1}{4 \pi} \ \overline{\alpha}_{\ k} 
\hspace*{0.2cm} ; \hspace*{0.2cm} 
\begin{array}{c}
d
 \vspace*{0.3cm} \\
\hline	\vspace*{-0.4cm} \\
d \ t 
\end{array}
\hspace*{0.2cm}
\overline{\kappa}_{\ k} \ =
\ - \ b_{\ k} \ ( \ \overline{\kappa}_{\ 1,2,3} \ ) \ \overline{\kappa}_{\ k}
\vspace*{0.3cm} \\
t \ = \ \log \ ( \ \mu^{\ 2} \ / \ m_{\ Z}^{\ 2} \ )
\hspace*{0.2cm} ; \hspace*{0.2cm} 
b_{\ k} \ = \ - \ \beta_{\ k} \ / \ \overline{g}_{\ k}
\vspace*{0.3cm} \\
b_{\ k} \ = \ B_{\ k} \ \overline{\kappa}_{\ k} \ + \ \cdots
\end{array}
\end{equation}

\noindent
We only illustrate the coupling constant evolution to one loop and ignore
the threshold effects above $m_{\ Z}$ caused by the interestingly
large value of the top quark mass ($\simeq 174 \ GeV$) :

\begin{equation}
\begin{array}{l}
\label{eq:48}
\left ( \ \overline{\alpha}_{\ k} \ \right )^{\ -1} 
\ \simeq 
\ \left ( \ \alpha_{\ k (0)} \ \right )^{\ -1} 
\ + \ \frac{1}{4 \pi} \ B_{\ k} \ t
\vspace*{0.3cm} \\
B_{\ 1} \ = \ - \ \frac{3}{5} \ \sum \ {\cal{Y}}^{\ 2}
\ \left \lbrace
\ \begin{array}{ll}
\frac{2}{3} & \mbox{chiral fermions} 
\vspace*{0.3cm} \\
\frac{1}{3} & \mbox{complex scalars} 
\end{array}
\right .
\vspace*{0.3cm} \\
B_{\ 2} \ = \ 7 \ \frac{1}{3} \ ( 6 )_{\ susy} \ - \ \sum_{\ doublets}
\ \left \lbrace
\ \begin{array}{ll}
\frac{1}{3} & \mbox{chiral fermions} 
\vspace*{0.3cm} \\
\frac{1}{6} & \mbox{complex scalars} 
\end{array}
\right .
\vspace*{0.3cm} \\
B_{\ 3} \ = \ 11 \ ( 9 \ )_{\ susy} \ - \ \sum_{\ 3 \ \& \ \overline{3}}
\ \left \lbrace
\ \begin{array}{ll}
\frac{1}{3} & \mbox{chiral fermions} 
\vspace*{0.3cm} \\
\frac{1}{6} & \mbox{complex scalars} 
\end{array}
\right .
\end{array}
\end{equation}

\noindent
In eq. (\ref{eq:48}) $\alpha_{\ k (0)}$ denote the values of the coupling constants at
scale $\mu \ = \ m_{\ Z}$ as given in eq. (\ref{eq:46}).

\noindent
We list the three constants $B_{\ k}$ for the standard model (SM) and 
the minimal supersymmetric standard model (MSSM) in table \ref{tabB} below
(abbreviating doublets by dblts).

\begin{table}[ht]
\[
\begin{array}{c@{\hspace*{0.6cm}}c@{\hspace*{0.4cm}}
c@{\hspace*{0.4cm}}c@{\hspace*{0.4cm}}c@{\hspace*{0.4cm}}
ccc} \\ \hline
 & \# \mbox{f dblts}  & \# \mbox{sc dblts} 
& \# \mbox{f 3 + $\overline{3}$} & \# \mbox{sc 3 + $\overline{3}$} 
& B_{\ 1} & B_{\ 2} & B_{\ 3}
\\ \hline
\mbox{SM} & 12 & 1 & 12 & 0 & - 4.1 & 3 \frac{1}{6} & 7
\vspace*{0.1cm}   \\
\mbox{MSSM} & 14 & 14 & 12 & 12 & - 6.6 & - 1 & 3
\vspace*{0.1cm} \\
\hline
\end{array}
\]
\caption{Matter fields in SM and MSSM.}
\label{tabB}
\end{table}

\noindent
The three pairs of coupling constants $k \ \leftrightarrow \ l$ with
$k \ \neq \ l$ cross at three respective crossing points

\begin{equation}
\begin{array}{l}
\label{eq:49}
t_{\ k l} \ = 
\begin{array}{c}
4 \pi
 \vspace*{0.3cm} \\
\hline	\vspace*{-0.4cm} \\
B_{\ l} \ - \ B_{\ k} 
\end{array}
\hspace*{0.2cm}
\left (
\left ( \ \alpha_{\ k (0)} \ \right )^{\ -1} 
\ - \ \left ( \ \alpha_{\ l (0)} \ \right )^{\ -1} 
\ \right )
\end{array}
\end{equation}

\begin{figure}[ht]
\begin{center}
\hspace*{-2cm}
\mbox{\epsfig{file=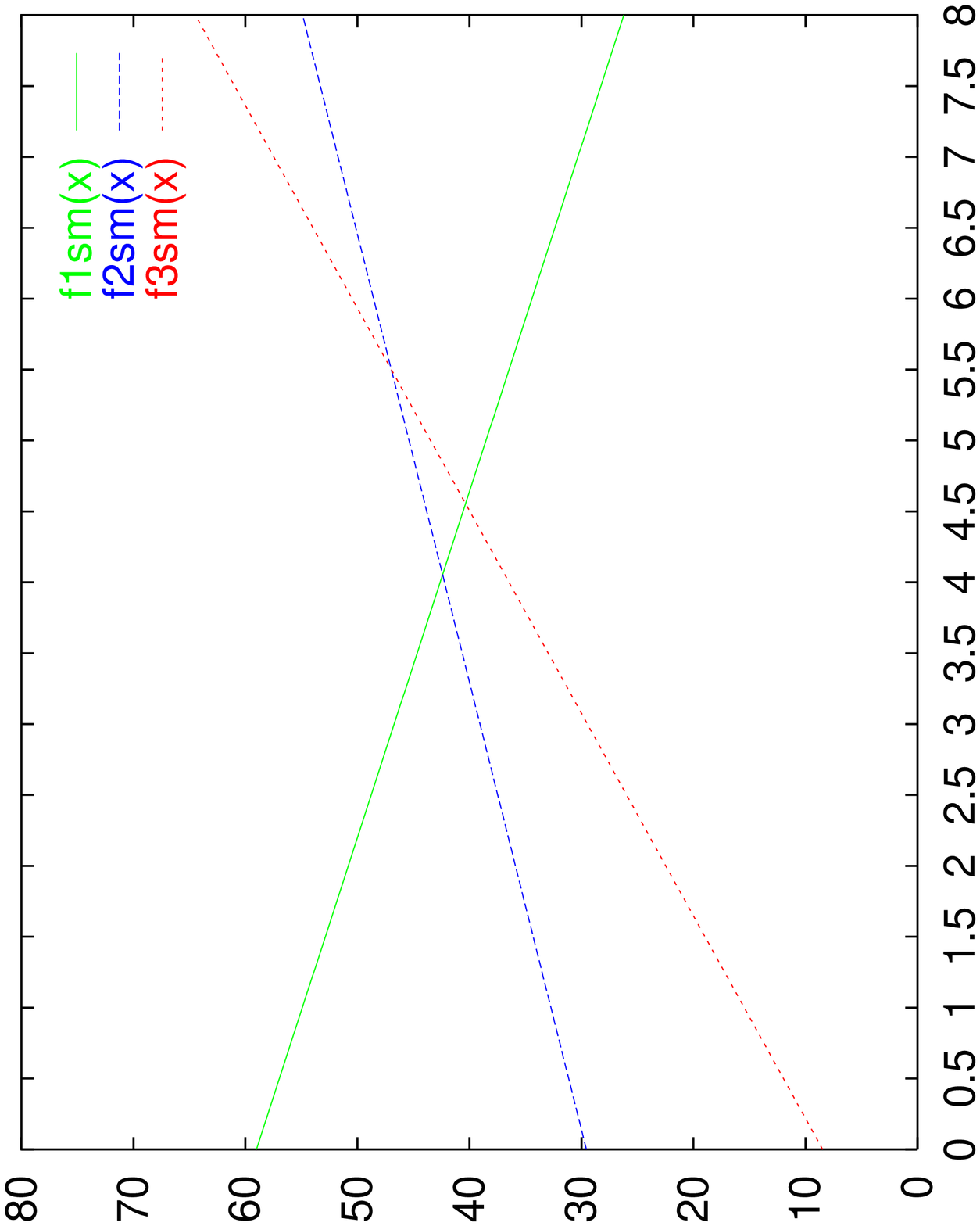,width=50mm}}
\hspace*{-3cm} (a) \hspace*{3cm}
\hskip 10mm
\mbox{\epsfig{file=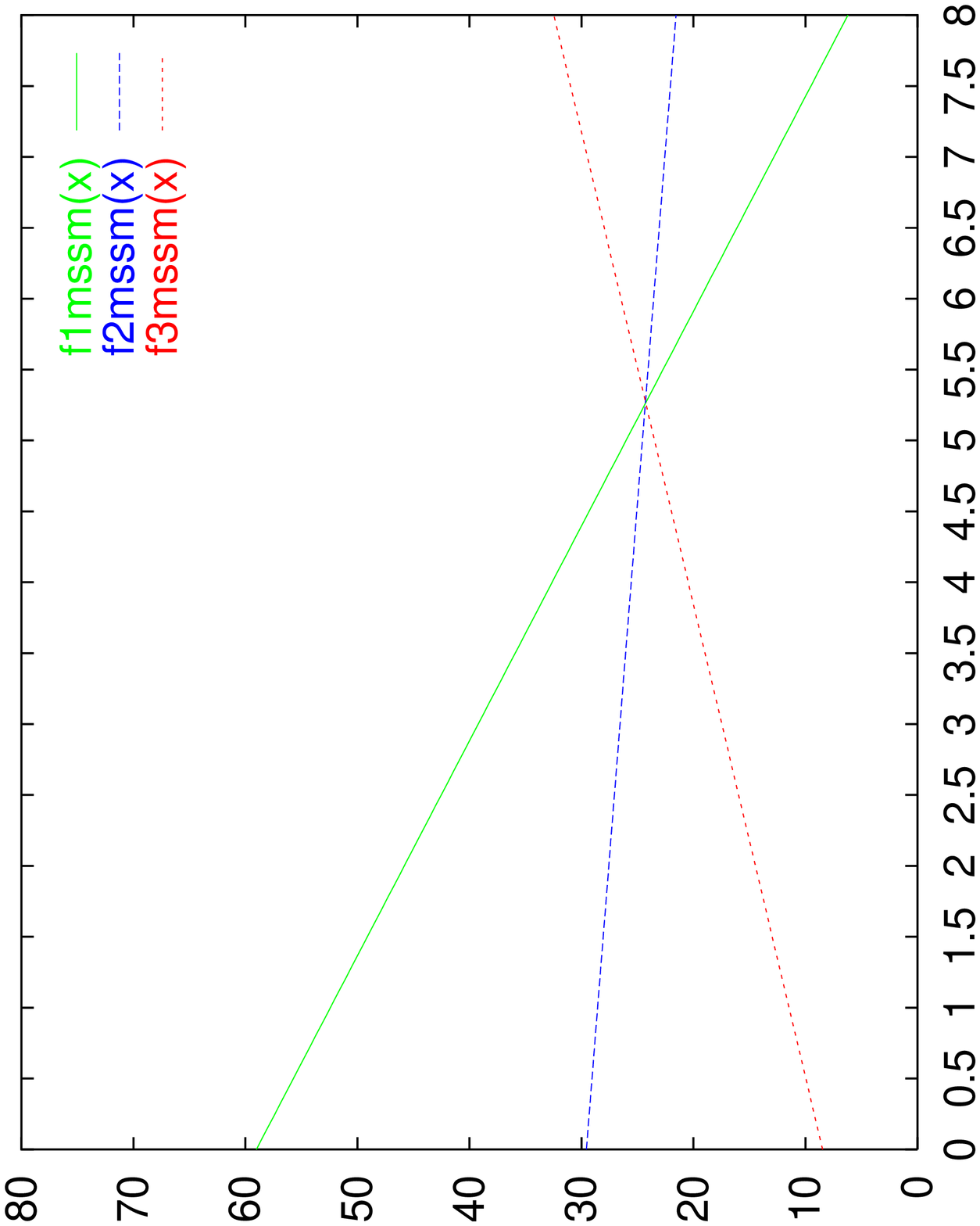,width=50mm}}
\hspace*{-3cm} (b)
\end{center}
\caption{
Coupling constant evolution according to SM (a) and MSSM (b)
field contents.
}
\label{smalpha}
\end{figure}

\noindent
The evolution of the (inverse) coupling constants according to the SM and the MSSM
\cite{LIGR}, \cite{zichnano} is shown
in figures \ref{smalpha} (a) and (b) respectively as a function of
$t \ = \ \log \ ( \ \mu^{\ 2} \ / \ m_{\ Z}^{\ 2} \ )$

\clearpage
\newpage

\noindent
The scales $M_{\ k l}$ corresponding to the crossing points 
$t_{\ k l} \ , \ k,l = 12, 13 , 23$ 
for the SM and the almost perfect unification scale $M_{\ susy}$ for MSSM are shown 
in table \ref{tabunif}

\begin{table}[ht]
\[
\begin{array}{c@{\hspace*{0.6cm}}c@{\hspace*{0.4cm}}
c@{\hspace*{0.4cm}}c} \\ \hline
M_{\ 1 2} \ (GeV) & M_{\ 13} \ (GeV) & M_{\ 2 3} \ (GeV) 
& M_{\ susy} \ (GeV) 
\\ \hline
1.05 \ 10^{\ 13} & 2.4 \ 10^{\ 14} & 0.9 \ 10^{\ 17} & 2.1 \ 10^{\ 16} 
\vspace*{0.1cm} \\
\hline
\end{array}
\]
\caption{SM crossing scales and MSSM unification scale.}
\label{tabunif}
\end{table}

\section*{7 + 8 SO10 and $\left ( \ \nu \ , \ {\cal{N}} \ \right )$ neutrino sector}
\vspace*{0.1cm}

\noindent
We focus here on the completion of the fermion families
to three 16 (spinor) representations of SO10 (spin10).

\noindent
The two decay chains of SO10 \cite{HFPM}, \cite{Georgi} 
proceed along SU5 and $SU4_{\ c}$ \cite{PaSa} as largest simple
subgroups

\begin{equation}
\begin{array}{l}
\label{eq:50}
\begin{array}{lll}
 & SO10 &
 \vspace*{0.3cm} \\
SU4_{\ c} \times SU2_{\ L} \times SU2_{\ R} & & SU5 \times U1_{\ {\cal{N}}}
 \vspace*{0.3cm} \\
SU3_{\ c} \times U1_{\ B-L} \times SU2_{\ R} & & SU3_{\ c} 
\times SU2_{\ L} \times U1_{\ {\cal{Y}}}  \times U1_{\ {\cal{N}}}
 \vspace*{0.3cm} \\
\multicolumn{3}{c} {SU3_{\ c} \times SU2_{\ L} \ \times U1_{\ {\cal{Y}}}} 
 \vspace*{0.3cm} \\
\multicolumn{3}{c} {SU3_{\ c} \times U1_{\ em}} 
\end{array}
\end{array}
\end{equation}

\noindent
The group $SU4_{\ c}$ in eq. (\ref{eq:50}) is the largest subgroup of
SO10 realized in a vectorlike fashion in the 16 spin10 representation attributed
to one family of (spin 1/2) fermions in the left chiral 4-d spinor basis :

\begin{equation}
\begin{array}{l}
\label{eq:51}
\frac{1}{2} \ \left ( \ 1 \ + \ \gamma_{\ 5}^{\ L} \ \right ) \ f \ \rightarrow 
\ f^{\ \dot{\gamma}}
 \vspace*{0.3cm} \\
f^{\ \dot{\gamma}} \ :
\ 16 \ =
\ \left (
 \begin{array}{ccc ccc c}
 4 & 2 & 1 & + & \overline{4} & 1 & 2
 \vspace*{0.3cm} \\
 SU4_{\ c} & SU2_{\ L} & SU2_{\ R} &  & SU4_{\ c} & SU2_{\ L} & SU2_{\ R}
\end{array}
\ \right )
 \vspace*{0.7cm} \\
\begin{array}{c ccc cc}
\updownarrow & SU2_{\ L}
& \multicolumn{2}{c} {\left (
\begin{array}{ccc c|cccc c}
u^{\ 1} & u^{\ 2} & u^{\ 3} & \nu_{\ e} \hspace*{0.2cm}  
& \hspace*{0.3cm} {\cal{N}}_{\ e} & \hat{u}^{\ 1} & \hat{u}^{\ 2} & \hat{u}^{\ 3} \\
 &  &  &  &  &  &  &  \\
d^{\ 1} & d^{\ 2} & d^{\ 3} & e^{\ -} \hspace*{0.2cm}
& \hspace*{0.3cm} e^{\ +} & \hat{d}^{\ 1} & \hat{d}^{\ 2} & \hat{d}^{\ 3}
\end{array}
\ \right )}
& SU2_{\ R} & \updownarrow 
 \vspace*{0.3cm} \\
 & & \hspace*{1.5cm} \longleftrightarrow &  \hspace*{0.5cm} \longleftrightarrow & &
 \vspace*{0.3cm} \\
 & & \hspace*{1.5cm} SU4_{\ c} & \hspace*{0.5cm} SU4_{\ c} & &
\end{array}
\end{array}
\end{equation}

\noindent
In the $SU5 \times U1_{\ {\cal{N}}}$ basis the 16 representation decomposes into 

\begin{equation}
\begin{array}{l}
\label{eq:52}
\left ( \ 1 \ , \ 5 \ \right )
\ + \ \left ( \ 10 \ , \ 1 \ \right )
\ + \ \left ( \ \overline{5} \ , \ - 3 \ \right )
 \vspace*{0.3cm} \\
\begin{array}{lll}
U1_{\ {\cal{N}}} & & SU5
 \vspace*{0.3cm} \\
5 &
\left (
\begin{array}{ccc c|cccc c}
 &  &  &  \hspace*{1.9cm}  
& \hspace*{0.3cm} {\cal{N}}_{\ e} &  &  & \hspace*{1.6cm} 
\end{array}
\ \right )
& 1
 \vspace*{0.5cm} \\
1 &
\left (
\begin{array}{ccc c|cccc c}
u^{\ 1} & u^{\ 2} & u^{\ 3} &  \hspace*{0.2cm}  
& \hspace*{0.3cm}  & \hat{u}^{\ 1} & \hat{u}^{\ 2} & \hat{u}^{\ 3} \\
 &  &  &  &  &  &  &  \\
d^{\ 1} & d^{\ 2} & d^{\ 3} &  \hspace*{0.2cm}
& \hspace*{0.3cm} e^{\ +} &  &  & 
\end{array}
\ \right )
& 10
 \vspace*{0.5cm} \\
- 3 &
\left (
\begin{array}{ccc c|cccc c}
 &  &  & \hspace*{1.05cm} \nu_{\ e} \hspace*{0.2cm} 
& \hspace*{0.3cm}  &  &  &  \\
 &  &  &  &  &  &  &  \\
 &  &  &  \hspace*{1.05cm} e^{\ -} \hspace*{0.2cm}
& \hspace*{0.6cm}  & \hat{d}^{\ 1} & \hat{d}^{\ 2} & \hat{d}^{\ 3} \hspace*{0.2cm}
\end{array}
\ \right )
& \overline{5}
\end{array}
 \vspace*{0.5cm} \\
\mbox{and :} 
\ \left (
\begin{array}{ll}
u & \nu_{\ e}
 \vspace*{0.3cm} \\
d & e^{\ -}
\end{array}
\ \right )
\ \rightarrow
\ \left (
\begin{array}{ll}
c & \nu_{\ \mu}
 \vspace*{0.3cm} \\
s & \mu^{\ -}
\end{array}
\ \right )
\ \rightarrow
\ \left (
\begin{array}{ll}
t & \nu_{\ \tau}
 \vspace*{0.3cm} \\
b & \tau^{\ -}
\end{array}
\ \right )
\end{array}
\end{equation}

\noindent
The integer eigenvalues of the $U1_{\ {\cal{N}}}$ generator in eq. (\ref{eq:52}) are normalized to

\begin{equation}
\begin{array}{l}
\label{eq:53}
\sum_{\ \left \lbrace 16 \right \rbrace}
\ \left ( \ U1_{\ {\cal{N}}} \ \right )^{\ 2} \ = \ 50
\end{array}
\end{equation}

\noindent
In order to achieve the same normalization as for $I_{\ 3 w}$ we would substitute

\begin{equation}
\begin{array}{l}
\label{eq:54}
U1_{\ {\cal{N}}} \ \rightarrow J_{\ {\cal{N}}}
\ = \sqrt{\frac{2}{50}} \ U1_{\ {\cal{N}}} 
\ = \frac{1}{5} \ U1_{\ {\cal{N}}} 
\end{array}
\end{equation}

\noindent
{\bf $\left ( \ \nu \ , \ {\cal{N}} \ \right )$ mass matrix}
 \vspace*{0.1cm} 

\noindent
The fields ${\cal{N}}_{\ e,\mu,\tau}$ are singlets under SU5 and a fortiori
under the SM gauge group. They can participate however in the so extended SM through
Yukawa interactions with the (or several) doublet scalars

\begin{equation}
\begin{array}{l}
\label{eq:55}
- \ {\cal{L}}_{\ Y} \ ( \ {\cal{N}} \ \ell \ \varphi^{\ *} \ )
\ = \ g^{\ (\nu,{\cal{N}})}_{\ I J}
\ {\cal{N}}^{\ I}_{\ \dot{\gamma}}
\ \left \lbrack 
\ \begin{array}{l}
\nu^{\ \dot{\gamma} \ J} \ \left ( \ \varphi^{\ 0} \ \right )^{\ *} \ +
 \vspace*{0.3cm} \\
\ell^{\ - \ \dot{\gamma} \ J}
\ \left ( \ \varphi^{\ -} \ \right )^{\ *}
\end{array}
\ \right \rbrack
\ + \ h.c.
 \vspace*{0.3cm} \\
 I \ , \ J \ = \ 1 \ , \ 2 \ , 3
\end{array}
\end{equation}

\noindent
The $u \ , \ c \ , t$-like Yukawa couplings defined in eq. (\ref{eq:55})
induce through the Higgs scalar vacuum expected value (eq. \ref{eq:40})
a Dirac mass matrix, pairing $\nu \ \leftrightarrow \ {\cal{N}}$, of the form 

\begin{equation}
\begin{array}{l}
\label{eq:56}
\left ( \ \mu \ = \ \frac{1}{\sqrt{2}} \ v 
\ g^{\ (\nu,{\cal{N}})} \ \right )_{\ I J}
\end{array}
\end{equation}

\noindent
The above notation of $\mu$ as the part of the neutrino mass matrix induced
by Yukawa couplings of the form given in eq. (\ref{eq:56}) shall not be confused
with the scale parameter $\mu$ used in the previous section.

\noindent
In addition an unrestricted Majorana mass term is compatible with the SM gauge group,
of the form

\begin{equation}
\begin{array}{l}
\label{eq:57}
- \ {\cal{L}}_{\ M} \ =
\ \frac{1}{2} \ M_{\ I J} 
\ {\cal{N}}^{\ I}_{\ \dot{\gamma}}
\ {\cal{N}}^{\ \dot{\gamma} \ J}
\ + \ h.c.
\hspace*{0.3cm} ; \hspace*{0.3cm} 
M_{\ I J} \ = \ M_{\ J I}
\end{array}
\end{equation}

\noindent
Apart from the symmetry condition for M in eq. (\ref{eq:57}) the three by three matrices
$\mu$ and M are (a priory) complex arbitrary and thus represent 18 + 12 real parameters.
They form the symmetric restricted six by six $\left ( \ \nu \ , \ {\cal{N}} \ \right )$
mass matrix 

\begin{equation}
\begin{array}{l}
\label{eq:58}
\vspace*{-0.3cm}
{\cal{M}} \ =
\begin{array}{l}
\vspace*{-1.4cm} \\
\ \begin{array}{ccc}
\hspace*{1.0cm} \nu & {\cal{N}} &
 \vspace*{0.3cm} \\
 \multicolumn{2}{c}
{\left (
\begin{array}{cc}
0 \hspace*{0.5cm} & \hspace*{0.5cm} \mu^{\ T}
 \vspace*{0.3cm} \\
\mu \hspace*{0.5cm} & \hspace*{0.5cm} M
\end{array}
\ \right )} 
& \begin{array}{c}
\nu
 \vspace*{0.3cm} \\
 {\cal{N}}
\end{array}
\end{array}
\end{array}
\end{array}
\end{equation}

\noindent
In eq. (\ref{eq:58}) $\mu^{\ T}$ denotes the transpose matrix of $\mu$.

\noindent
B - L invariance, with $L \ ( \ {\cal{N}} \ ) \ = \ - 1$, is violated 
by the Majorana mass term at some larege scale.

\noindent
For details of mass and mixing structure see e.g. ref. \cite{CHPM}.

\noindent
The 0 entry in ${\cal{M}}$ gives rise to one exact mass relation, valid through renormalization.
This involves the 3 nonnegative eigenvalues of the hermitian three by three matrix
$\mu \ \mu^{\ \dagger}$ and the six physical neutrino masses of
the full (hermitian) mass matrix ${\cal{M}} \ {\cal{M}}^{\ \dagger}$. The latter
are assumed to contain three light and three heavy masses :

\begin{equation}
\begin{array}{l}
\label{eq:59}
\mu \ \mu^{\ \dagger} \ \rightarrow \ \mu_{\ 1,2,3} \ \ge 0
 \vspace*{0.3cm} \\
{\cal{M}} \ {\cal{M}}^{\ \dagger} \ \rightarrow \ m_{\ 1,2,3} \ , \ M_{\ 1,2,3} \ \ge \ 0
\hspace*{0.3cm} ; \hspace*{0.3cm} 
M_{\ J} \ \gg \ m_{\ K}
\end{array}
\end{equation}

\noindent
The relation which is known by the name sea-saw \cite{MGM}, \cite{Yanag} becomes 

\begin{equation}
\begin{array}{l}
\label{eq:60}
Det_{\ 6 \times 6} \ {\cal{M}} \ = 
\ \left ( \ Det_{\ 3 \times 3} \ \mu \ \right )^{\ 2} 
\hspace*{0.3cm} \rightarrow \hspace*{0.3cm} 
m_{\ 1/3} \ M_{\ 1/3} \ =
\ \left ( \ \mu_{\ 1/3} \ \right )^{\ 2}
 \vspace*{0.3cm} \\
m_{\ 1/3} \ =
\ \left ( 
\ m_{\ 1} \ m_{\ 2} \ m_{\ 3} 
\ \right )^{\ 1/3}
\hspace*{0.3cm} ; \hspace*{0.3cm} 
M_{\ 1/3} \ =
\ \left ( 
\ M_{\ 1} \ M_{\ 2} \ M_{\ 3} 
\ \right )^{\ 1/3}
 \vspace*{0.3cm} \\
\mu_{\ 1/3} \ =
\ \left ( 
\ \mu_{\ 1} \ \mu_{\ 2} \ \mu_{\ 3} 
\ \right )^{\ 1/3}
\end{array}
\end{equation}

\noindent
A generic {\it estimate} can be obtained assuming at unification scale
the SO10 inspired relation of equal mass matrices for the $( \ \nu \ , \ N \ )$ Yukawa
couplings and for u,c,t quarks and values of the light neutrino
masses characteristic of the mass square differences compatible with
the solar neutrino deficit and the atmospheric neutrino anomaly
\cite{BaDa}, \cite{JBa}, \cite{SupK} 

\begin{equation}
\begin{array}{l}
\label{eq:61}
\overline{\mu}_{\ unif} \ = \ \overline{\mu}_{\ u,c,t}
\hspace*{0.3cm} \rightarrow \hspace*{0.3cm} 
\mu_{\ 1/3} \ \simeq 
\ \frac{1}{3} \ \left ( \ m_{\ u} \ m_{\ c} \ m_{\ t} \ \right )^{\ 1/3}
\ \simeq \ 0.4 \ GeV
 \vspace*{0.3cm} \\
 m_{\ 1} \ \simeq \ m_{\ 2} \ \simeq \ 10^{\ -3} \ eV
\hspace*{0.3cm} ; \hspace*{0.3cm} 
 m_{\ 3} \ \simeq \ 5 \ 10^{\ -2} \ eV
 \vspace*{0.3cm} \\
\rightarrow \  m_{\ 1/3} \ \simeq \ 3.6 \ 10^{\ -3} \ eV
\hspace*{0.3cm} \Rightarrow \hspace*{0.3cm} 
M_{\ 1/3} \ \simeq \ 4.4 \ 10^{\ 11} \ GeV
\end{array}
\end{equation}

\noindent
The order of magnitude estimate in eq. (\ref{eq:61})
$ M_{\ 1/3} \ \simeq \ 4.4 \ 10^{\ 11} \ GeV$ reveals a high mass scale which
serves through inverse powers as a protection of lepton number(s)
as well as B-L approximate conservation at low energy. 
Nevertheless this mass scale is several
orders of magnitude smaller than the unification scale obtained from the MSSMoin table
\ref{tabunif}.

\noindent
Thus the successful neutrino oscillation interpretation of the associated solar
and atmospheric anomalies \cite{BaDa}, \cite{JBa}, \cite{SupK} is, since
this workshop took place, augmented by the hint of the observation of a light Higgs boson
with a mass $\simeq 115 \ GeV$ at LEP \cite{aleph}, \cite{L3} and 
by the measurement of the anomalous magnetic moment of the muon at BNL
\cite{gmuon}, \cite{gmuonth} as yet uncertain but characteristic signatures
of small deviations from SM expectations. They all point towards intermediary mass scales
between the electroweak scale as characterized by $v$ or $m_{\ Z}$ and the
unification scale for charges , represented within the MSSM by 
$M_{\ susy} \ \simeq \ 2 \ 10^{\ 16} \ GeV$, bearing on physics beyond the
standard model.

\noindent
One aspect related to spontaneous generation of mass in QCD for (the main mass of) nucleons and 
through the Higgs effect in the SM or its supersymmetric extensions for all elementary quanta
except neutrinos, was the central topic in our discussion of the cosmological
constant in sections 1-6 as summarized in table \ref{tabcosw}. This topic has an 
apparently paradoxical yet important bearing
on the unification of charge like and spin like gauge interactions, including 
gravity among the latter.

\noindent
I would like to appreciate the outstanding contributions of Alberto Sirlin
to the backbones of the perturbative structure of gauge field theory. They were
for me guidelines since student years and they serve to substancify that
the quest for unification is not a dream but based on physical and logical reality - - for all I know.
\vspace*{0.1cm}

\section*{9 + 10 Outlook ( quo vadis ? )}
\vspace*{0.1cm}

\noindent
\begin{enumerate}
\item Spontaneous susy breaking in flat 4-d space time generates  vacuum energy density

\begin{displaymath}
\begin{array}{l}
\left \langle \ \Omega \ \right | \ \vartheta_{\ \mu \nu}
\ \left | \ \Omega \ \right \rangle \ = \ \varepsilon_{\ susy} \ g_{\ \mu \nu}
\hspace*{0.3cm} \mbox{with} \hspace*{0.3cm}
\varepsilon_{\ susy} \ > \ 0
\end{array}
\end{displaymath}

This sign is opposite to the corresponding vacuum energy densities in 
QCD and with respect to the Higgs effect.

\item The metric $g_{\ \mu \nu}$ does not react through d=4 curvature 
to any of the vacuum energy densities 
$\varepsilon_{\ susy} \ > \ 0$, $\varepsilon_{\ QCD, Higgs effect} \ < \ 0$.
Through the absorption of goldstino modes into gravitinos
the mass scale of susy braking is linked to gravitino masses, which at present
remain unknown.

\item This absence of 4-d curvature ($AdS_{\ 4} \ \leftrightarrow \ \varepsilon < 0$ ,
$dS_{\ 4} \ \leftrightarrow \ \varepsilon  >  0$) indicates indirectly the presence of extra
dimensions with unknown curvature scale.

\item All ungauged U1-like symmetries : B, $L_{\ e,\mu,\tau}$, B-L $\cdots$
are broken and protected, except for B,  at low energy by the mass of heavy neutrino
flavors ${\cal{N}}$ : $M_{\ 1/3} \ ( {\cal{N}} ) \ \sim \ 10^{\ 11} \ GeV$.

\item While signatures of the path towards unification ( of all gauge and associated
symmetries) appear -- to me -- to be elusive as to the scale of $\sim \ 1$ TeV,
they hopefully will manifest themselves first in small effects or the absence of expected
large ones and this rightly so will demand a definite alertness in spirit
and a wide open mind towards consistent speculations.
\end{enumerate}


\newpage

\begin{thebibliography}{99}
  
\bibitem{BraDi} C.Brans and R. H. Dicke, {\it Phys. Rev.} {\bf 124} (1961) 925. 

\bibitem{Connes} A. Connes, 'Noncommutative geometry: year 2000' ,
e-print archive: math.qa/0011193. 

\bibitem{ShiVaZa} M. Shifman, A. Vainshtain and V. Zakharov,
		  {\it Nucl. Phys.} {\bf B147} (1979) 385. 

\bibitem{Nari} S. Narison, {\it Nucl. Phys.} {\bf B509} (1998) 312.

\bibitem{SKPM} S. Kabana and P. Minkowski, 
               'Mapping out the QCD phase transition in multiparticle production',
	       e-print archive: hep-ph/0010247. 

\bibitem{Riess} A. G. Riess et al., {\it Astron. J.} {\bf 116} (1998) 1009,
		e-print archive : astro-ph/9805201.

\bibitem{Perlm} S. Perlmutter et al., {\it Astrophys. J.} {\bf 517} (1999) 565.
		e-print archive : astro-ph/0812133.

\bibitem{Dok} Yu.I. Dokshitzer, V.A. Khoze, A.H. Mueller and S.I. Troyan,
              'Basics of perturbative QCD' , Editions Fronti\'{e}res,
	       Gif-sur-Yvette, France, 1991.

\bibitem{leut} H. Leutwyler, 'Non-lattice determinations of the light quark masses' ,
               talk given at Lattice 2000, Bangalore, India, Aug. 2000, 
               e-print archive : hep-ph/0011049. 

\bibitem{PMCz} P. Minkowski, {\it Czech. J. Phys.} {\bf B 40} (1990) 1003.

\bibitem{Higgs} P.W. Higgs, {\it Phys.Lett.} {\bf 12} (1964) 132         
		and 'Spontaneous symmetry breakdown without massless bosons' ,
                {\it Phys.Rev.} {\bf 145} (1966) 1156.

\bibitem{EngBr} F. Englert and R. Brout, 
		'Broken symmetry and the mass of gauge vector mesons' ,
                {\it Phys.Rev.Lett.} {\bf 13} (1964) 321. 

\bibitem{JEDR} J. Ellis and D.Ross, 'A light Higgs Boson would invite Supersymmmetry',
               e-print archive : hep-ph/0012067. 

\bibitem{JEllis} J. Ellis, 'Perspectives in High-Energy Physics' ,
                 e-print archive : hep-ph/0007161. 

\bibitem{LBPM} L. Bergamin and P. Minkowski, 
	       'Spontaneous susy breaking in N=2 super Yang-Mills theories' ,
               e-print archive : hep-ph/0011041.

\bibitem{WM} W. J. Marciano, these proceedings.

\bibitem{LIGR} L.E. Ibanez, G.G. Ross, {\it Phys.Lett.} {\bf B105} (1981) 439.  

\bibitem{zichnano} J.L. Lopez, D.V. Nanopoulos and A. Zichichi,
           'The superworlds of SU(5) and SU(5) X U(1): a critical assessment and overview',
            in proceedings of International School of Subnuclear Physics: 30th Course:
            'From Superstrings to the Real Superworld', Erice, Italy, 14-22 Jul 1992, 
            Erice Subnuclear 1992:0311-3.

\bibitem{HFPM} H. Fritzsch and P. Minkowski, {\it Ann. Phys.} {\bf 93} (1975) 193.

\bibitem{Georgi} H. Georgi, {\it AIP Conf.Proc.} {\bf 23} (1975) 575, (No.2).

\bibitem{PaSa} J. Pati and A. Salam, {\it Phys. Rev.} {\bf D10} (1974) 275.

\bibitem{CHPM} C. Heusch and P. Minkowski, {\it Int. J. Mod. Phys..} {\bf A15} (2000) 2429.

\bibitem{MGM} M. Gell-Mann, P. Ramond and R. Slansky  in {\it Supergravity},
	      D. Freedman and P. van Nieuwenhuizen eds., North Holland 1979.

\bibitem{Yanag} T. Yanagida, {\it Proc. Workshop on Unified Theory and
	      Baryon Number in the Universe}, O. Sawada and A. Sugamoto eds., KEK 1979.

\bibitem{BaDa} J. Bahcall and R. Davis, 'The evolution of neutrino astronomy',
	       e-print archive : astro-ph/9911486.

\bibitem{JBa} J. Bahcall, these proceedings.

\bibitem{SupK} 'Neutrino-induced upward stopping muons in Super-Kamiokande',
               The Super-Kamiokande Collaboration, {\it Phys. Lett.} {\bf B467} (1999) 185.

\bibitem{aleph} ALEPH Collaboration (R. Barate et al.), 
         'Observation of an excess in the search for the standard model Higgs boson at ALEPH',
          {\it Phys.Lett.} {\bf B495} (2000) 1, e-print archive: hep-ex/0011045. 

\bibitem{L3} L3 Collaboration (M. Acciarri et al.), 
          'Higgs candidates in e+ e- interactions at s**(1/2)-GeV,
          {\it Phys.Lett.} {\bf B495} (2000) 18, e-print archive: hep-ex/0011043. 

\bibitem{gmuon} Muon g-2 Collaboration (H.N. Brown et al.), 
         'Precise measurement of the positive muon anomalous magnetic moment', 
         e-print archive: hep-ex/0102017. 

\bibitem{gmuonth} A. Czarnecki and W. J. Marciano, 
           'The muon anomalous magnetic moment: a harbinger for "new physics"'.
           e-print archive: hep-ph/0102122. 

\end{thebibliography}
\end{document}